\documentclass[journal,natbib,screen,cspaper,transmag]{IEEEtran}

\usepackage{booktabs} 
\usepackage[square,numbers]{natbib}
\usepackage{multicol}
\usepackage{multirow}
\usepackage[inline]{enumitem}
\usepackage[skip=0pt]{caption}
\usepackage{acronym}
\usepackage{amssymb,amsthm,amsmath}
\usepackage{subcaption}
\usepackage{graphicx}
\usepackage{url}
\usepackage{tikz,xcolor,hyperref}

\acrodef{RNN}{recurrent neural network}
\acrodef{SR}{sequential recommendation}
\acrodef{SAC-SR}{shared account cross-domain sequential recommendation}
\acrodef{PSJNet}{Parallel Split-Join Network}
\acrodef{GRU}{gated recurrent unit}
\acrodef{MLP}{multilayer perceptron}
\acrodef{RBM}{restricted Boltzmann machines}
\acrodef{CF}{collaborative filtering}
\acrodef{KNN}{K-nearest neighbors}
\acrodef{LDA}{linear Discriminant analysis}
\acrodef{LFM}{latent factor model}
\acrodef{ReLU}{rectified linear unit}
\acrodef{BPTT}{back-propogation through time}
\acrodef{MDP}{Markov decision process}
\acrodef{MDPs}{Markov decision processes}
\acrodef{DSSM}{deep structured semantic model}
\acrodef{MC}{Markov chain}
\acrodef{MF}{matrix factorization}

\DeclareMathOperator{\softmax}{softmax}

\AtBeginDocument{%
  \providecommand\BibTeX{{%
    \normalfont B\kern-0.5em{\scshape i\kern-0.25em b}\kern-0.8em\TeX}}}

\IEEEpubid{0000--0000/00~\copyright~2021 IEEE}

\begin{document}

\title{Parallel Split-Join Networks for Shared Account Cross-domain Sequential Recommendations
\thanks{$^*$Corresponding author.}
\thanks{This paper is a substantially extended version of~\citep{ma-2019-pi-net}. The additions are three-fold. 
First, we unify the parallel modeling framework introduced in~\citep{ma-2019-pi-net} into the \ac{PSJNet} architecture introduced in this paper and propose a new model \ac{PSJNet}-\uppercase\expandafter{\romannumeral2} that improves the performance over previous proposals ($\pi$-Net in~\citep{ma-2019-pi-net} corresponds to \ac{PSJNet}-\uppercase\expandafter{\romannumeral1} in this paper). 
Second, we build a new dataset for \acl{SAC-SR} by simulating shared account characteristics on a public dataset. 
Third, we carry out more experiments to test \ac{PSJNet}-\uppercase\expandafter{\romannumeral1} and  \ac{PSJNet}-\uppercase\expandafter{\romannumeral2}. 
More than half of the experiments reported in this paper were not in \citep{ma-2019-pi-net} and all relevant result tables and figures are either new additions to the article or report new results.}}


\definecolor{lime}{HTML}{A6CE39}
\DeclareRobustCommand{\orcidicon}{%
    \begin{tikzpicture}
    \draw[lime, fill=lime] (0,0) 
    circle [radius=0.16] 
    node[white] {{\fontfamily{qag}\selectfont \tiny ID}};    \draw[white, fill=white] (-0.0625,0.095) 
    circle [radius=0.007];    \end{tikzpicture}
    \hspace{-2mm}}
\foreach \x in {A, ..., Z}{%
    \expandafter\xdef\csname orcid\x\endcsname{\noexpand\href{https://orcid.org/\csname orcidauthor\x\endcsname}{\noexpand\orcidicon}}
    }

\newcommand{\orcidauthorG}{0000-0002-1086-0202}

\author{Wenchao Sun, Muyang Ma, Pengjie Ren$^*$, Yujie Lin, Zhumin Chen, Zhaochun Ren, Jun Ma, Maarten de Rijke\orcidG
}

\IEEEtitleabstractindextext{%

\begin{abstract}
\Acl{SR} is a task in which one models and uses sequential information about user behavior for recommendation purposes.
We study \acl{SR} in a particularly challenging context, in which multiple individual users share a single account (i.e., they have a shared account) and in which user behavior is available in multiple domains (i.e., recommendations are cross-domain).
These two characteristics bring new challenges on top of those of the traditional \acl{SR} task.
First, we need to identify the behavior associated with different users and different user roles under the same account in order to recommend the right item to the right user role at the right time.
Second, we need to identify behavior in one domain that might be helpful to improve recommendations in other domains.

In this work, we study \emph{\acl{SAC-SR}} and propose \textbf{P}arallel \textbf{S}plit-\textbf{J}oin \textbf{Net}work (\ac{PSJNet}), a parallel modeling network to address the two challenges above. 
We use ``split'' to address the challenge raised by shared accounts; \ac{PSJNet} learns role-specific representations and uses a gating mechanism to filter out, from mixed user behavior, information of user roles that might be useful for another domain.
In addition, ``join'' is used to address the challenge raised by the cross-domain setting; \ac{PSJNet} learns cross-domain representations by combining the information from ``split'' and then transforms it to another domain.
We present two variants of \ac{PSJNet}, \ac{PSJNet}-\uppercase\expandafter{\romannumeral1} and \ac{PSJNet}-\uppercase\expandafter{\romannumeral2}.
\ac{PSJNet}-\uppercase\expandafter{\romannumeral1} is a ``split-by-join'' framework that splits the mixed representations to get role-specific representations and joins them to obtain cross-domain representations at each timestamp simultaneously.
\ac{PSJNet}-\uppercase\expandafter{\romannumeral2} is a ``split-and-join'' framework that first splits role-specific representations at each timestamp, and then the representations from all timestamps and all roles are joined to obtain cross-domain representations.
We concatenate the in-domain and cross-domain representations to compute a recommendation score for each item.
Both \ac{PSJNet}-\uppercase\expandafter{\romannumeral1} and \ac{PSJNet}-\uppercase\expandafter{\romannumeral2} can simultaneously generate recommendations for two domains where user behavior in two domains is synchronously shared at each timestamp.

We use two datasets to assess the effectiveness of \ac{PSJNet}.
The first dataset is a simulated \acl{SAC-SR} dataset obtained by randomly merging the Amazon logs from different users in the movie and book domains.
The second dataset is a real-world \acl{SAC-SR} dataset built from smart TV watching logs of a commercial organization.
Our experimental results demonstrate that \ac{PSJNet} outperforms state-of-the-art \acl{SR} baselines in terms of MRR and Recall.
\end{abstract}

}
\maketitle
\IEEEdisplaynontitleabstractindextext

\acresetall
\parskip0pt


\section{Introduction}

It is hard to apply traditional recommendation methods such as, e.g., \ac{CF}-based  methods~\cite{Sarwar:2001:ICF:371920.372071} or \ac{MF}-based models~\cite{Koren:2009:MFT:1608565.1608614,8540438}, in recommendation scenarios in which user profiles may be absent.
This happens when users are not logged in or the recommender system does not track user-ids.
For this reason, \acfi{SR} has been proposed for session-based recommendations~\cite{no:19}.
Compared with traditional recommendations, \ac{SR} has natural advantages when it comes to sequential dynamics~\cite{he2016fusing}, i.e., \ac{SR} methods may generate different lists of recommended items at different timestamps.
The goal of \ac{SR} is to generate recommendations based on a sequence of user behavior (e.g., a sequence of songs listened to, videos watched, or products clicked), where interactions are usually grouped by same time frame~\cite{no:3,no:25,cheng2017exploiting}.
\acp{SR} have a wide range of applications in many domains such as e-commerce, job websites, music and video recommendations~\cite{no:4}.
\IEEEpubidadjcol
And users usually have specific goals during the process, such as finding a good restaurant in a city, or listening to a song of a certain style or mood~\citep{no:1}.

Early studies into \ac{SR} are mostly based on \acp{MC}~\cite{zimdars2001using} or \ac{MDPs}\acused{MDP}~\cite{no:4}. 
Sequences of items are considered as states and a state-transition matrix or function is learned to generate recommendations.
In this way, the dynamic characteristics of \ac{SR} are taken into account.
However, because the states in a \ac{MC}- or \ac{MDP}-based method correspond to sequences of items, the state-transition matrix or function quickly becomes unmanageable in realistic scenarios~\cite{Quadrana:2018:SRS:3236632.3190616}.
Recurrent neural networks have demonstrated their effectiveness in sequence modeling in the field of natural language processing.
Motivated by this, recent studies have introduced \acp{RNN} into \ac{SR} \cite{no:19} and today \ac{RNN}-based models have been widely adopted in the context of \ac{SR}.
Various \ac{RNN} architectures have been proposed to enhance \ac{SR}, e.g., to make \acp{SR} context-aware~\cite{7837948}, personalized~\cite{no:1}, or deal with repeat behavior~\cite{no:48}.
However, so far \ac{RNN}-based methods focus on a single domain and none simultaneously considers the shared account and cross-domain scenarios.

\begin{figure*}[ht]
\centering
\includegraphics[width=0.65\textwidth]{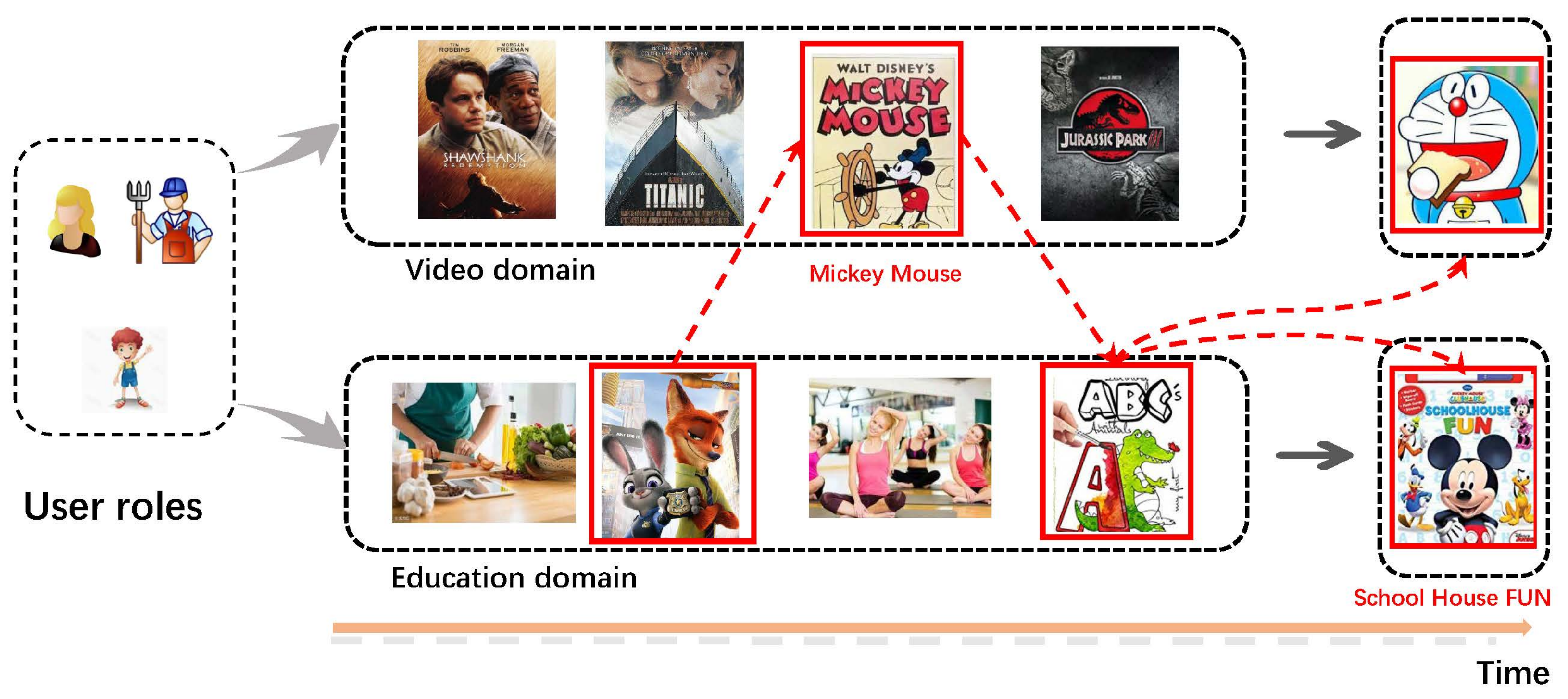}
\caption{The smart TV scenario provides a natural example of \acfi{SAC-SR}. Here, members of a family (the ``user roles'') share the same account. Moreover, the \emph{video} domain contains various movies, TV series, cartoons, etc. The \emph{education} domain contains educational programs and technical tutorials, etc. \boxed{Boxed} items reflect similar user interests. Red lines show the interactions and connections between user behavior in the two domains.}
\label{01-1}
\end{figure*}

In this paper, we study \ac{SR} in a particularly challenging context, \acfi{SAC-SR}.
In this context multiple individual users share a single account (i.e., they have a \emph{shared account}) and user behavior is recorded in multiple domains (i.e., recommendations should be \emph{cross-domain}).
The shared account task is considered because in some recommendation applications people tend to share a single account, resulting in multiple ``user roles'' under each account.
For example, in the smart TV recommendation scenario depicted in Figure~\ref{01-1}, members of a family correspond to different user roles, e.g., ``parents'', ``children'', and they share a single account to watch videos.
The existence of shared accounts makes it harder to generate relevant recommendations, because the behavior of multiple user roles is mixed together.
Note that user roles do not explicitly represent specific users.
We consider user roles instead of users because the number of user roles is smaller than that of users, and it is generally easier to distinguish different user roles than users.
We consider the cross-domain task because it is a common phenomenon in practice:
users use different platforms to access domain-specific services in daily life.
We can get user behavior data from different domains during the same time period.
For example, many smart TV platforms use different channels to provide different services, e.g., a video channel (domain) that offers movies or TV series and an educational channel that offers educational material, as depicted in Figure~\ref{01-1}.
User behavior in one domain may be helpful for improving recommendations in another domain~\cite{no:35,no:39,no:42} because user behavior in different domains may reflect similar user interests.
For example, as illustrated in Figure~\ref{01-1}, videos like ``Mickey Mouse'' in the video domain might help to predict the next item ``School House Fun'' in the educational domain because they reflect the same interest in the Disney cartoon character ``Mickey Mouse'' presumably by a child in this family.
Although leveraging user behavior information from another domain may provide useful information to help improve the recommendation performance, this type of transfer is non-trivial because the behavior of multiple user roles is mixed and this may introduce noise.
This raises an interesting challenge, namely how to identify behavior from one domain that might be helpful to improve recommendations in other domains while minimizing the impact of noisy information?

In prior work that focuses on shared accounts, a common approach is to capture user interests by extracting latent features from high-dimensional spaces that describe the relationships among user roles under the same account \cite{no:52,no:57,no:56}.
And in prior work on the cross-domain task, one common solution is to aggregate information from two domains~\cite{no:30,no:33,no:34}, while another is to transfer knowledge from the source domain to the target domain~\cite{no:43,no:50}. 
None of these methods can be directly applied to \acs{SAC-SR}: either important sequential characteristics of \ac{SR} are largely ignored or they rely on explicit user ratings, which are usually unavailable in \ac{SR}.
In our previous work~\cite{ma-2019-pi-net}, we have introduced an architecture ($\pi$-Net) that addresses the above issues by simultaneously generating recommendations for two domains where user behavior from two domains is synchronously shared at each timestamp.

In this work, we generalize over the $\pi$-Net architecture with a more general framework, namely the \textbf{P}arallel \textbf{S}plit-\textbf{J}oin \textbf{Net}work (\ac{PSJNet}), that introduces the ``split'' and ``join'' concepts to address the shared account and cross-domain characteristics in \ac{SAC-SR}.
To address shared accounts, ``split'' is used to identify behavior of different user roles, where we employ a gating mechanism to extract role-specific representations containing information of some user roles that might be useful for another domain from mixed user behavior.
To address the cross-domain aspect, ``join'' is used to discriminate and combine useful user behavior; we learn cross-domain representations by combining the information from ``split'' and then adopting it to another domain.

Specifically, \ac{PSJNet} is organized in four main modules, namely a \textit{sequence encoder}, a \textit{split unit}, a \textit{join unit}, and a \textit{hybrid recommendation decoder}.
The \textit{sequence encoder} module encodes the current sequence of mixed user behavior from each domain into a sequence of in-domain representations.
Then, depending on how ``split'' and ``join'' are implemented, we present two \ac{PSJNet} variants, i.e., \ac{PSJNet}-\uppercase\expandafter{\romannumeral1} and \ac{PSJNet}-\uppercase\expandafter{\romannumeral2}.
\ac{PSJNet}-\uppercase\expandafter{\romannumeral1} (which corresponds to $\pi$-Net) employs a ``Split-by-Join'' scheme where it splits the mixed representations to get role-specific representations and joins them to get cross-domain representations at each timestamp simultaneously.
\ac{PSJNet}-\uppercase\expandafter{\romannumeral2} employs a ``Split-and-Join'' scheme where it first splits role-specific representations at each timestamp, and then the representations from all timestamps and all roles are joined to obtain cross-domain representations.
For both variants, ``split'' and ``join'' are operated in a parallel recurrent way, which means that they can synchronously share information across both domains at each timestamp.
Finally, the \textit{hybrid recommendation decoder} module estimates the recommendation scores for each item based on the information from both domains, i.e., the in-domain representations from the target domain and the cross-domain representations from the complementary domain.
During learning, \ac{PSJNet} is jointly trained on two domains in an end-to-end fashion.

To assess the effectiveness of \ac{PSJNet}, we need datasets that exhibit both shared account and cross-domain characteristics.
To the best of our knowledge, there is no such real-world dataset that is publicly available.
We construct two datasets for \acs{SAC-SR}.
The first dataset is a simulated \acs{SAC-SR} dataset obtained by randomly merging the logs from different users in the movie and book domains from a well-known Amazon dataset.\footnote{\url{http://jmcauley.ucsd.edu/data/amazon/}}
Although the dataset can satisfy experimental requirements, the merged user behavior is not realistic and is unlikely to reflect realistic scenarios.
Therefore, we build a second dataset from smart TV watching logs of a commercial company, which is a real-world \acs{SAC-SR} scenario.
We release both datasets to facilitate future research.
We carry out extensive experiments on both datasets. 
The experimental results show that \ac{PSJNet} outperforms state-of-the-art baselines in terms of MRR and Recall. 
We also conduct ablation studies to show that the proposed parallel ``split'' and ``join'' schemes are effective and useful for \acs{SAC-SR}.

The additional contributions of this paper compared to our previous work in \cite{ma-2019-pi-net} are as follows:
\begin{itemize}[leftmargin=*,nosep]
\item We present the \ac{PSJNet} framework, which introduces the ``split'' and ``join'' concepts to address the shared account and cross-domain characteristics of \ac{SAC-SR}.
\item We reformulate the previous proposal $\pi$-Net as \ac{PSJNet}-\uppercase\expandafter{\romannumeral1} within the \ac{PSJNet} framework, and propose a new variant \ac{PSJNet}-\uppercase\expandafter{\romannumeral2} that further improves the performance. 
\item We carry out experiments on two datasets for \ac{SAC-SR}. One is constructed by simulating shared account characteristics on a public dataset, the other is a real-world dataset. And we conduct additional experiments on these two datasets to show the effectiveness of the two \ac{PSJNet} variants.
\end{itemize}


\section{Related work}

We consider three types of related work: sequential recommendations, shared account recommendations, and cross-domain recommendations.

\subsection{Sequential recommendations}

It is hard to capture sequential dynamics in recommendation scenarios with classical recommendation methods such as \acf{MF}- or \acf{CF}-based methods.
Instead, dedicated methods have been developed for \ac{SR} or next basket recommendation.

\subsubsection{Traditional methods} 
The traditional approaches for \ac{SR} are mostly based on \acfp{MC}~\cite{zimdars2001using} or \acf{MDPs}~\cite{no:4} to predict a user's next action given their last action~\cite{no:16}. 
\citet{zimdars2001using} are the first to propose \acp{MC} for web page recommendation. 
They investigate how to extract sequential patterns to learn the next state using probabilistic decision-tree models. 
To further improve the performance, \citet{no:4} propose an \ac{MDP}-based recommendation method, where the next recommendation can be computed through the transition probabilities among items.
To combine the advantages of \ac{MF} and \ac{MC}-based methods, \citet{Rendle:2010:FPM:1772690.1772773} propose a method based on personalized transition graphs over an underlying \ac{MC}.
They show that the proposed method subsumes both a common \ac{MC} and the normal \ac{MF} model.
\citet{yap2012effective} introduce a competence score measure in personalized sequential pattern mining for next-item recommendations. 
\citet{chen2012playlist} take playlists as an \ac{MC}, and propose logistic Markov embeddings to learn representations of songs for playlists prediction. 
\citet{Wu:2013:PNR:2507157.2507215} propose Personalized Markov Embedding (PME) to consider sequential singing behavior for the next song recommendation.
They embed users and songs into a Euclidean space, where the distance between songs and users represent their relationships.
Given a user's last song, they can generate personalized recommendations by ranking the candidate songs according to the relationships.

\citet{doi:10.1137/1.9781611972832.71} argue that source domain data is not always consistent with the observations in the target domain, which may misguide the target domain recommendation. 
They propose a criterion based on empirical prediction error and its variance to better capture the consistency across domains in \ac{CF} settings. 
To address the sparsity and long-tailed distribution issues of most recommendation datasets and take sequential dynamics into consideration at the same time, \citet{he2016fusing} propose to combine the advantages of \ac{MC}-based methods and \ac{CF}-based methods.
They fuse a similarity-based method with \ac{MC} to learn a personalized weighting scheme over the sequence of items to characterize users in terms of both interests and the strength of sequential behavior.

All of the \ac{MC} or \ac{MDP}-based sequential recommendation methods mentioned above show improvements by modeling sequential dynamics.
However, a major issue they share is that they can only consider a very short sequence (e.g., the most recent five items in \citep{no:4}), because the state space quickly becomes unmanageable when taking long sequences into account~\cite{Quadrana:2018:SRS:3236632.3190616}.

\subsubsection{Deep learning-based methods}
Recently, \acfp{RNN} have been devised to model variable-length sequential data~\cite{no:18,10.1145/3426723}.
\citet{no:19} are the first to apply \acp{RNN} to sequential recommendation and achieve significant improvements over traditional methods. 
They utilize session-parallel mini-batch training and employ ranking-based loss functions to train the recommendation model. 
In later work, they propose data augmentation techniques to improve the performance of \acp{RNN}~\cite{no:20}.

Contextual information has proved to be very important for behavior modeling in traditional recommendations. 
\citet{7837948} incorporate contextual information into \ac{SR} and propose a context-aware \ac{RNN} model. 
Instead of using the constant input matrix and transition matrix from conventional RNN models, their CA-RNN employs adaptive matrices.
The authors use context-specific input matrices to capture external conditions under which user behavior happens, such as time, location, weather and so on. 
They also use context-specific transition matrices to capture how the length of time intervals between adjacent behavior in historical sequences affects the transition of global sequential features.
\citet{no:22} investigate how to add item property information such as text and images to an \ac{RNN} framework and introduce a number of parallel \ac{RNN} architectures (p-\acp{RNN}); they propose alternative training strategies for p-\acp{RNN} that suit them better than standard training. 
\citet{no:21} explore user's dwell time based on an existing \ac{RNN}-based framework by boosting items above a predefined dwell time threshold. 
\citet{8534409} incorporate visual and textual information and propose MV-\ac{RNN} to alleviate the cold start problem.

\citet{Donkers:2017:SUR:3109859.3109877} introduce a new gated architecture with additional input layers for \acp{GRU} to explicitly represent individual users, for the purpose
of generating personalized next item recommendations.
\citet{9319527} propose a dictionary learning-based approach to model a user's static and dynamic preferences. They use a \ac{GRU} to translate a user's sequential behavior into dynamic user preferences.
\citet{unknown} propose a dynamic memory-based recurrent attention network for modeling long behavior sequences.
\citet{no:1} propose a hierarchical \ac{RNN} model that can be used to generate personalized sequential recommendations. 
\citet{no:25} explore a hybrid encoder with an attention mechanism to model the user's sequential behavior and intent to capture the user's main purpose in the current sequence. 
\citet{no:50} propose a novelty seeking model based on sequences in multi-domains to model an individual's propensity by transferring novelty seeking traits learned from a source domain for improving the accuracy of recommendations in the target domain. 
\citet{Tang:2018:PTS:3159652.3159656} propose a convolutional sequence embedding recommendation model for personalized top-n sequential recommendation to address the more recent items where they argue that more recent items in a sequence have a larger impact on the next item.
\citet{no:48} propose a repeat-aware \ac{RNN} model to address the repeat consumption in \ac{SR}, which is a common phenomenon in many recommendation scenarios where the same item is repeatedly re-consumed.
They incorporate a new repeat recommendation mechanism into \acp{RNN} that can choose items from a user's history and recommends them at the right time. 

Memory enhanced \acp{RNN} have been well studied for \ac{SR} recently.
\citet{Chen:2018:SRU:3159652.3159668} argue that existing \ac{SR} methods usually embed a user's historical records into a single latent representation, which may have lost the per item- or feature-level correlations between a user's historical records and future interests.
They introduce a memory mechanism to \ac{SR} and design a memory-augmented neural network integrated with insights from collaborative filtering.

\citet{Huang:2018:ISR:3209978.3210017} propose a knowledge enhanced \ac{SR} model to capture fine-grained user interests from interaction sequences. 
They integrate \ac{RNN}-based networks with a key-value memory network. 
They further incorporate knowledge base information to enhance the learned semantic representations.
\citet{Ma:2018:MRM:3209978.3210026} propose a cross-attention memory network to perform the mention recommendation task for multi-modal tweets where they make full use of both textual and visual information.
\citet{Huang:2019:TMR:3289600.3290972} introduce a taxonomy-aware multi-hop reasoning network, which integrates a basic \ac{GRU}-based sequential recommender with an elaborately designed memory-based multi-hop reasoning architecture. 
They incorporate taxonomy data as structural knowledge to enhance the reasoning capacity. 

\citet{Wang:3209978.3210026} hypothesize that the collaborative information contained in neighborhood sequences (that have been generated previously by other users and reflect similar user intents as the current sequence) might help to improve recommendation performance for the current sequence.
They propose a \ac{RNN} model with two parallel memory modules: one to model a user's own information in the current sequence and the other to exploit collaborative information in neighborhood sequences~\cite{no:24}.
Most recently, \citet{9319534} propose a transformer-based structured intent-aware model that first extracts intents from sequential contexts, and then adopts an intent graph to capture the correlations among user intents.

\subsection{Shared account recommendations}
Most recommender systems assume that every account in the data represents a single user. 
However, multiple users often share a single account. 
A typical example is a smart TV account for the whole family. 

Previous approaches to shared account recommendations typically first identify users and then make personalized recommendations \cite{no:52,no:56,no:43,no:57}.
\citet{no:55} are the first to study user identification, in which they use linear subspace clustering algorithms; they recommend the union of items that are most likely to be rated highly by each user. 
\citet{no:51} propose a similarity-based channel clustering method to group similar channels for IPTV accounts, and they use the Apriori algorithm to separate users that are merged under a single account. 
After that, they use personal profiles to recommend additional channels to the account. 
\citet{no:53} assume that different users consume services in different periods.
They decompose users based on mining different interests over different time periods from consumption logs. 
Finally, they use a standard User-KNN method to generate recommendations for each identified user. 
\citet{no:54} also analyze the similarity of the proportion of each type of items within a time period to judge whether a sequence is generated by the same user. 
Then, they generate recommendations by recommending personalized genres to the identified users. 
\citet{7837939} develop a time-aware user identification model based on Latent Dirichlet Allocation using a hidden variable to represent the user, and assume consumption times to be generated by latent time topics. 
\citet{yang2017personalized} identify users by using a projection based unsupervised method, and then use Factorization Machine techniques to predict a user's interest based on historical information to generate personalized recommendations. 
\citet{no:49} propose an unsupervised learning-based framework to identify users and differentiate the interests of users and group sessions by users. 
They construct a heterogeneous graph to represent items and use a normalized random walk to create item-based session embeddings. 
A hybrid recommender is then proposed that linearly combines the account-level and user-level item recommendation by employing Bayesian personalized ranking matrix factorization~\cite{no:61}.


\subsection{Cross-domain recommendations}

Cross-domain recommendation concerns data from multiple domains, which has proven to be helpful for alleviating the cold start problem~\cite{no:26,no:27} and data sparsity issues~\cite{no:28,no:29}. 
There is an assumption that there exists overlap in information between users and/or items across different domains \cite{no:30,no:31}.

\subsubsection{Traditional methods} 
There are two main ways for dealing with cross-domain recommendations~\cite{no:32}.
One is to aggregate knowledge between two domains. 
\citet{no:37} propose four mediation techniques to solve the data sparsity problem by merging user interests and extracting common attributes of users and items. 
\citet{Tang:2012:CCR:2339530.2339730} propose a cross-domain topic learning model to address three challenges in cross-domain collaboration recommendation: sparse connections (cross-domain collaborations are rare); complementary expertise (cross-domain collaborators often have different expertise and interest) and topic skewness (cross-domain collaboration topics are focused on a subset of topics)
\citet{no:39} compare several collaborative methods to demonstrate the effectiveness of utilizing available preference data from Facebook. 
\citet{no:38} model user interests by using \ac{MF} separately on different domains, and then incorporate user interaction patterns that are specific to particular types of items to generate recommendations on the target domain. 
\citet{8742537} propose to discover both explicit and implicit similarities from latent factors across domains based on matrix tri-factorization.
\citet{no:40} propose a consensus regularization classifier framework by considering both local data available in source domain and the prediction consensus with the classifiers learned from other source domains. 
\citet{no:41} construct a nonparametric Bayesian framework by jointly considering multiple heterogeneous link prediction tasks between users and different types of items. 
\citet{no:44} exploit the users and items shared between domains as a bridge to link different domains by embedding all users and items into a low-dimensional latent space between different domains. 
\citet{9346486} utilize both \ac{MF} and an attention mechanism for fine-grained modeling of user preferences; the overlapping cross-domain user features are combined through feature fusion.

The other approach to cross-domain recommendation is to transfer knowledge from the source domain to the target domain.
\citet{no:42} propose tensor-based factorization to share latent features between different domains.
\citet{no:43} propose a code-book-transfer to transfer rating patterns between domains.
\citet{no:46} propose a content-based approach to learn semantic information between domains.
\citet{9356204} propose a transition-based cross-domain collaborative filtering method to capture both within- and between-domain transitions of user feedback sequences.
\citet{9360540} propose a method that not only transfers item's learned latent factors, but also selectively transfers user's latent factors.


\subsubsection{Deep learning-based methods}
Deep learning is well suited to transfer learning as it can learn high-level abstractions among different domains~\citep{onal-neural-2018}.
\citet{no:36} introduce a factorization framework to tie collaborative filtering and content-based filtering together; they use neural networks to build user and item embeddings. 
\citet{no:30} propose a multi-view deep learning recommendation system by using rich auxiliary features to represent users from different domains based on \acf{DSSM} \cite{huang2013learning}. 
\citet{Fernandez-Tobias:2016:ADC:2959100.2959175} address the cold-start issue in a target domain by exploiting user interests from a related auxiliary domain.
They show that cross-domain information is useful to provide more accurate and diverse recommendations when user feedback in the target domain is scarce or not available at all. 
\citet{no:34} propose a model using a cross-stitch network \cite{misra2016cross} to learn complex user-item interaction relationships based on neural collaborative filtering \cite{no:33}. 
\citet{no:50} propose a novelty-seeking model to fully characterize users' novelty-seeking trait so as to obtain a better performance across domains. 
\citet{no:35} are the first to introduce the problem of cross-domain social recommendations; they combine user-item interactions in information domains (such as online travel planning) and user-user connections in social network services (such as Facebook or Twitter) to recommend relevant items of information domains to target users of social domains; user and item attributes are leveraged to strengthen the embedding learning.  

Although the methods proposed in the studies listed above have been proven to be effective in many applications, they either cannot be applied to sequential recommendations or do not consider the shared account or cross-domain characteristics.
In our previous work, we have proposed $\pi$-Net in order to address shared account and cross-domain characteristics in sequential recommendations by extracting information of different user roles under the same account and transferring it to a complementary domain at each timestamp~\citep{ma-2019-pi-net}.
In this work, we present a more general framework called \ac{PSJNet}: $\pi$-Net can be  viewed as a particular instantiation of \ac{PSJNet} and we propose another instantiation that further improves the recommendation performance over $\pi$-Net.


\section{Method}
In this section, we first provide a formulation of the \acf{SAC-SR} problem. 
Then, we introduce \ac{PSJNet} and describe two instantiations of the framework.
For each variant, we give a high-level introduction and describe each component in detail.

\subsection{Shared account cross-domain sequential recommendation}
We represent a cross-domain behavior sequence (e.g., watching videos, reading books) from a shared account as $S=\langle A_1,B_1,B_2,\ldots, A_i, \ldots, B_j, \ldots\rangle$, 
where $A_i \in \mathbb{A} $ $(1 \leq  i \leq N)$ is the index of a single consumed item in domain $A$;
$\mathbb{A}$ is the set of all items in domain $A$;
$B_j \in \mathbb{B}$ $(1 \leq  j \leq M)$ is the index of a single consumed item in domain $B$;
$\mathbb{B}$ is the set of all items in domain $B$;
$N$ and $M$ are the number of items in the sequences from domain $A$ and $B$, respectively.
Given $S$, \acs{SAC-SR} tries to predict the next item by computing the recommendation probabilities for all candidates in two domains respectively, as shown in Eq.~\ref{equation1}:
\begin{equation}
\label{equation1}
\begin{split}
&P(A_{i+1}\mid S) \sim f_A(S) \\
&P(B_{j+1}\mid S) \sim f_B(S), \\
\end{split}
\end{equation}
where $P(A_{i+1}\mid S)$ denotes the probability of recommending the item $A_{i+1}$ that will be consumed next in domain $A$.
Also, $f_A(S)$ is the model or function used to estimate $P(A_{i+1}\mid S)$.
Similar definitions apply to $P(B_{j+1}\mid S)$ and $f_B(S)$.

The main differences between \acs{SAC-SR} and traditional \ac{SR} are two-fold.
First, in \acs{SAC-SR}, $S$ is generated by multiple users (e.g., family members) while it is usually generated by a single user in \ac{SR}.
Second, \acs{SAC-SR} considers information from both domains for the particular recommendations in one domain, i.e., $S$ is a mixture of behavior from multiple domains.
In this paper, we only consider two domains but the ideas easily generalize to multiple domains.

Next, we will describe two \ac{PSJNet} variants in detail.
The key idea of \ac{PSJNet} is to learn a recommendation model that can first extract the information of some specific user roles from domain $A$, and then transfer the information to domain $B$, and combine it with the original information from domain $B$ to improve recommendation performance for domain $B$, and vice versa. 
This process is achieved in a parallel way, which means that the information from both domains is shared recurrently. 

\subsection{Sequence encoder}
\label{section:sequence-encoder}
Both variants of \ac{PSJNet} that we consider use the same sequence encoder.
Like previous studies \cite{no:19,no:20,no:1}, we use a \ac{RNN} to encode a sequence $S$. 
Here, we employ two separate \acp{GRU} as the recurrent units to encode the items from domain $A$ and domain $B$ respectively.
And the \ac{GRU} is given as follows:
\begin{equation}
\begin{split}
&z_t=\sigma(W_z[emb(x_t),h_{t-1}])  \\
&r_t=\sigma(W_r[emb(x_t),h_{t-1}])  \\
&\widetilde{h_t}=\tanh(W_{\widetilde{h}}[emb(x_t),r_t\odot h_{t-1}]) \\
&h_t=(1-z_t)\odot h_{t-1} + z_t\odot \widetilde{h_t},
\end{split}
\end{equation}
where $W_z$, $W_r$, and $W_{\widetilde{h}}$ are weight matrices;  $emb(x_t)$ is the item embedding of item $x$ at timestamp $t$; and $\sigma$ denotes the sigmoid function.
The initial state of the \acp{GRU} is set to zero vectors, i.e., $h_0=0$.
Through the \textit{sequence encoder} we obtain $H_A=\langle h_{A_1}$, $h_{A_2}$, \ldots, $h_{A_i}$, \ldots, $h_{A_N}\rangle$ for domain $A$, and $H_B=\langle h_{B_1},h_{B_2},\ldots,h_{B_j},\ldots,h_{B_M}\rangle$ for domain $B$. 
We consider the last state as the in-domain representation, i.e., $h_A = h_{A_N}$ for domain $A$ and $h_B = h_{B_M}$ for domain $B$.
The in-domain representations are combined with the cross-domain representations (i.e., $h_{(A\rightarrow B)}$ or $h_{(B\rightarrow A)}$) to compute the final recommendation score.
In the next two subsections, we will describe two \ac{PSJNet} instantiations that adopt different frameworks to learn the cross-domain representations.


\subsection{\ac{PSJNet}-\uppercase\expandafter{\romannumeral1}}
\label{section:pi-i-walkthrough}

\begin{figure*}[htb]
\centering
\includegraphics[width=.65\textwidth]{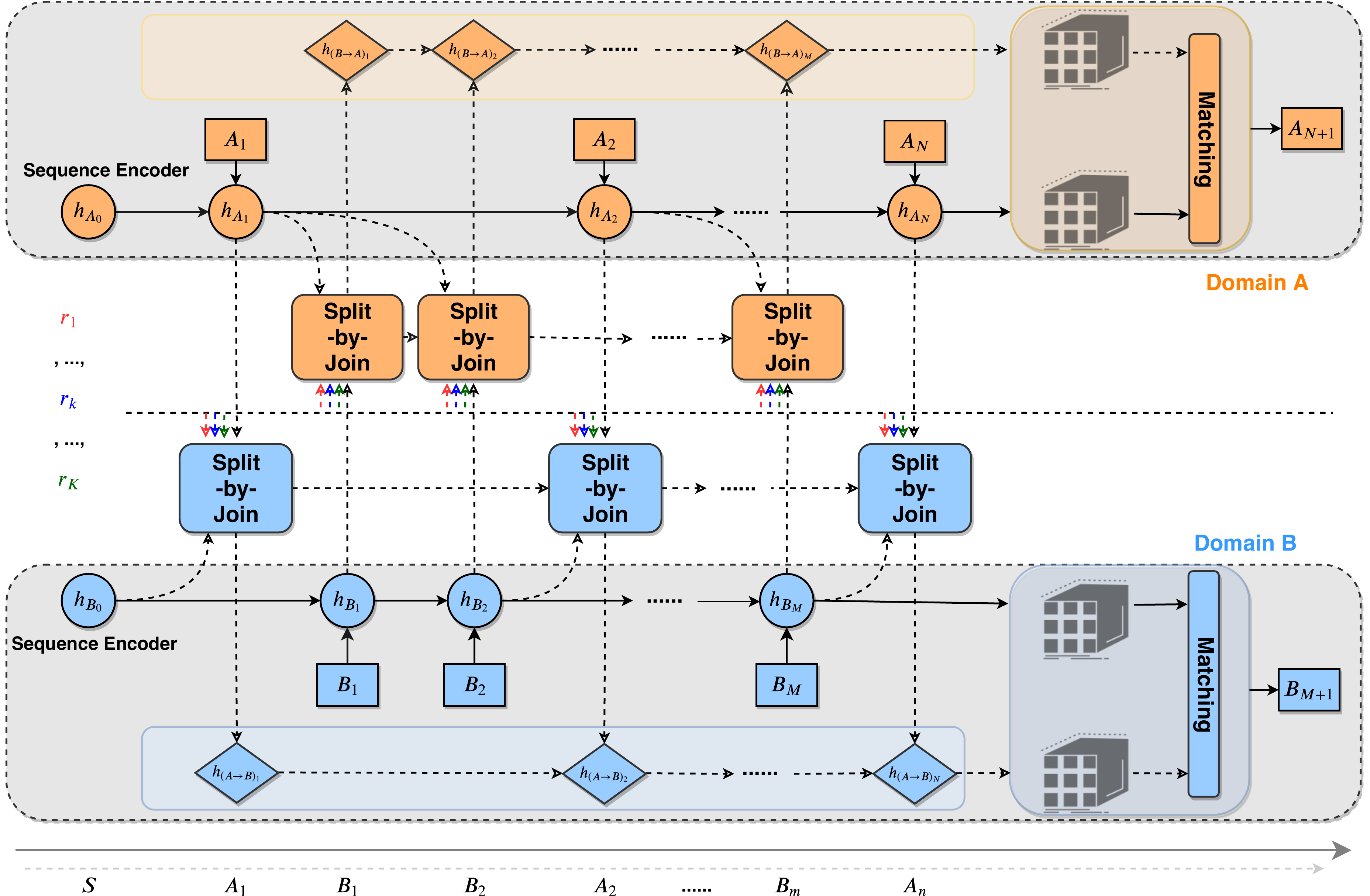}
\vspace*{1mm}
\caption{An overview of \ac{PSJNet}-\uppercase\expandafter{\romannumeral1}. 
The orange and blue colors represent different domains.
Red, purple and green represent different user roles. Section~\ref{section:pi-i-walkthrough} contains a walkthrough of the model.}
\label{02}
\end{figure*}

In this subsection, we describe \ac{PSJNet}-\uppercase\expandafter{\romannumeral1} in detail.
\ac{PSJNet}-\uppercase\expandafter{\romannumeral1} is a reformulation of $\pi$-Net~\cite{ma-2019-pi-net} within the \ac{PSJNet} framework, where we reformulate the shared account filter unit (SFU) and the cross-domain transfer unit (CTU) as a split-by-join unit.
Figure~\ref{02} provides an overview of \ac{PSJNet}-\uppercase\expandafter{\romannumeral1}.
\ac{PSJNet}-\uppercase\expandafter{\romannumeral1} is a ``Split-by-Join'' framework where it gets the role-specific representations from the mixed user behavior and simultaneously joins them at each timestamp.
Then the representations are transformed to another domain as cross-domain representations.
\ac{PSJNet}-\uppercase\expandafter{\romannumeral1} consists of three main components: 
\begin{enumerate*}[label=(\arabic*)]
\item a \textit{sequence encoder} (see Section~\ref{section:sequence-encoder}), 
\item a \textit{split-by-join unit}, and 
\item a \textit{hybrid recommendation decoder} (see Section~\ref{section:recommendation-decoder}). 
\end{enumerate*}
The \textit{sequence encoder} encodes the behavior sequences of each domain into high-dimensional hidden representations. 
The \textit{split-by-join unit} takes each domain's representations as input and tries to first split the representations of specific user roles, and then joins and transforms them to another domain at each timestamp $t$.
The \textit{matching decoder} combines the information from both domains and generates a list of recommended items.
Please refer to Sections~\ref{section:sequence-encoder} and~\ref{section:recommendation-decoder} for details of the \textit{sequence encoder} and the \textit{hybrid recommendation decoder}, respectively.
In this subsection, we focus on the core module (i.e., the \textit{split-by-join unit}) of \ac{PSJNet}-\uppercase\expandafter{\romannumeral1}.

\subsubsection{Split-by-join unit}
Under the shared account scenario, the behavior recorded under the same account is generated by different user roles. 
In practice, not all user roles that share the account are active in all domains.
Besides, even though some user roles are active in the same domain, they may have different interests.
For example, in the smart TV scenario, children may occupy the majority of the educational channel, while adults dominate the video TV channel.

\begin{figure*}
\centering
\includegraphics[width=.65\textwidth]{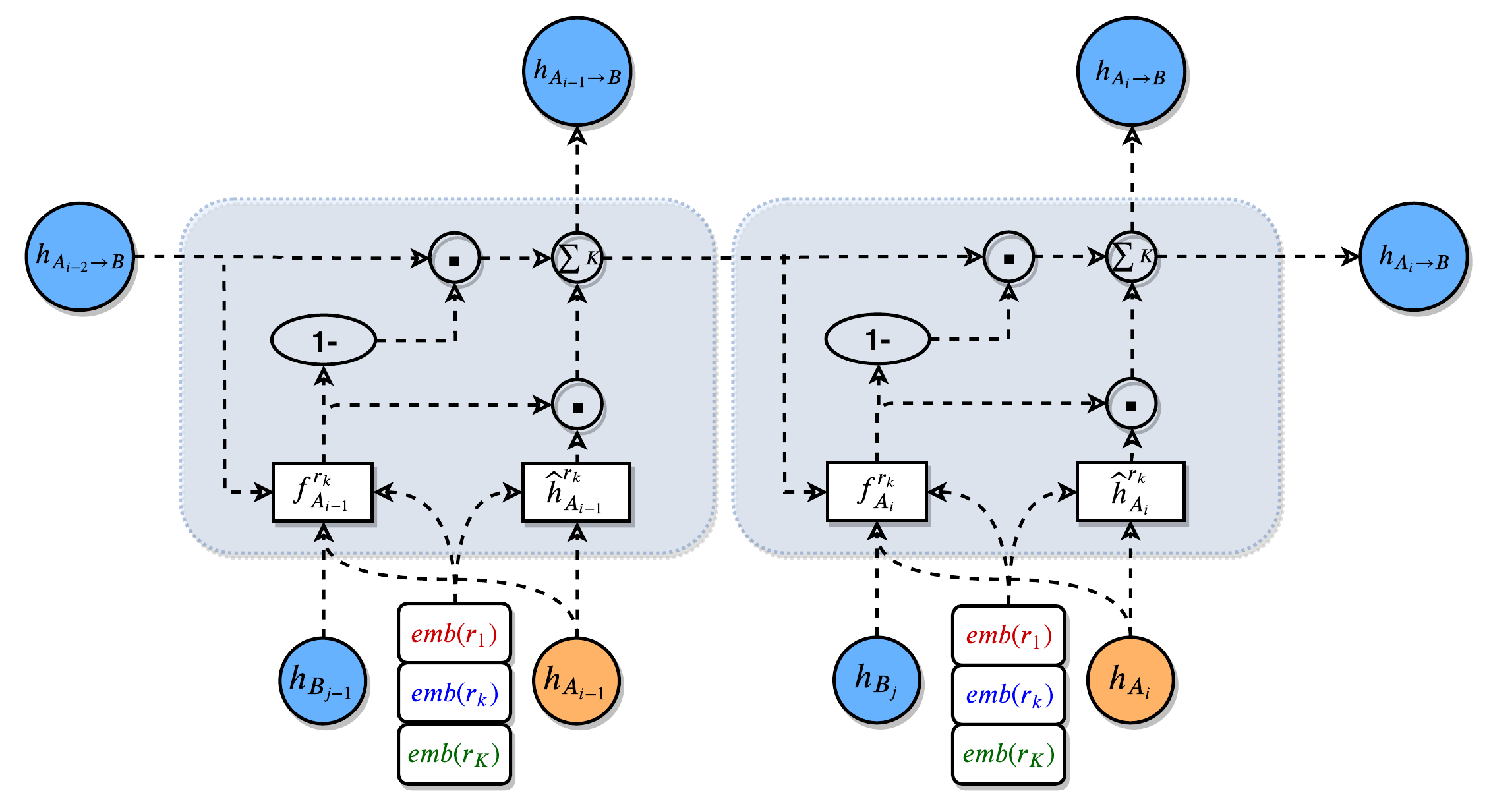}
\caption{The \textit{split-by-join unit} illustrated while transforming information from domain $A$ to domain $B$.}
\label{03}
\end{figure*}

The outputs $H_A$ or $H_B$ of the \textit{sequence encoder} are the mixed representations of all user roles sharing the same account.
To learn role-specific representations from these mixed representations, we propose a \textit{split-by-join unit}, as shown in Figure~\ref{03}.
The \textit{split-by-join unit} can be applied bidirectionally from ``domain $A$ to domain $B$'' and ``domain $B$ to domain $A$,'' meaning that the information is extracted from one domain and transferred to the other domain.
Here, we take the ``domain $A$ to domain $B$'' direction and achieving recommendations in domain $B$ as an example.
To learn role-specific representations, we need to know the number of user roles under each account, which is, unfortunately, unavailable in most cases.
In this work, we assume that there are $K$ latent roles ($r_1$, $r_2$, \ldots , $r_{k}$, \ldots, $r_{K}$) under each account.
For example, in a family account, the user roles may correspond to child, male parent, female parent, etc.
We first embed each latent role into $emb(r_k)$ $(1 \leq  k \leq K)$, which represents and encodes the latent interests of each role.
Then, we split the specific representation for role $r_k$ at timestamp $i$ in domain $A$ with Eq.~\ref{equation3}:
\begin{equation}
\label{equation3}
h_{A_i}^{r_k}= f_{A_i}^{r_k}\odot \widehat{h}_{A_i}^{r_k} + (1-f_{A_i}^{r_k})\odot h_{A_{i-1}\rightarrow B},
\end{equation}
where $\odot$ denotes element-wise multiplication.
Mathematically, the representation $h_{A_i}^{r_k}$ is a combination of two representations $\widehat{h}_{A_i}^{r_k}$ and $h_{A_{i-1}\rightarrow B}$ balanced by $f_{A_i}^{r_k}$.
A higher value of $f_{A_i}^{r_k}$ means that item $A_i$ is more likely generated by $r_k$ and we should pay more attention to $r_k$'s related representation $\widehat{h}_{A_i}^{r_k}$.
A lower value of lower $f_{A_i}^{r_k}$ means that item $A_i$ might not be related to $r_k$ and we should inherit more information from previous time steps. 

Next, we introduce the definitions of the three parts of Eq.~\ref{equation3} one by one.

\begin{enumerate}[label=(\alph*),leftmargin=*]
\item We propose a gating mechanism to implement $f_{A_i}^{r_k}$ in Eq.~\ref{equation4}:
\begin{equation}
\label{equation4}
\begin{split}
f_{A_i}^{r_k}=  \sigma (W_{f_A} \cdot h_{A_i} +{}& W_{f_B} \cdot h_{B_j} + U_f \cdot  h_{A_{i-1}\rightarrow B} \\& + V_f \cdot emb(r_k) + b_f),
\end{split}
\end{equation}
where $\cdot$ means matrix multiplication; 
$W_{f_A}$, $W_{f_B}$, $U_f$ and $V_f$ are the parameters; $b_f$ is the bias term; $h_{A_i}$ and $h_{B_j}$ are the mixed representations of domain $A$ and $B$ at timestamp $i$ and $j$, respectively.
Note that $B_j$ is the last item from domain $B$ before $A_i$ in the mixed sequence.
$h_{A_{i-1}\rightarrow B}$ is the previous output, which will be explained later (under item~\ref{item-c}).
After the sigmoid function $\sigma$, each value of $f_{A_i}^{r_k}$ falls into $(0, 1)$.
Thus, the gating score $f_{A_i}^{r_k}$ controls the amount of information of $r_k$ to transfer from domain $A$ to domain $B$.
It should be noted that each latent representation $emb(r_k)$ indicates the distribution of user roles with similar interests under each account, and it does not explicitly represents a specific user.

\item 
$\widehat{h}_{A_i}^{r_k}$ is the candidate representation for $r_k$ at timestamp $i$ in domain $A$, which is computed based on the mixed representation $h_{A_i}$, the filtered previous output $h_{A_{i-1}\rightarrow B}$, and the user role $r_k$'s latent interest $emb(r_k)$, as shown in Eq.~\ref{equation5}:
\begin{equation}
\label{equation5}
\begin{split}
\widehat{h}_{A_i}^{r_k}=\tanh (&W_h \cdot h_{A_i} + U_h \cdot h_{A_{i-1}\rightarrow B} +{} \\& V_h \cdot emb(r_k) + b_h ), 
\end{split}
\end{equation}
where $W_h$, $U_h$ and $V_h$ are the parameters; $b_h$ is the bias term.

\item \label{item-c}
$h_{A_{i}\rightarrow B}$ is the final output of the cross-domain representation at timestamp $i$ from domain $A$ to domain $B$, 
which is calculated by joining each role-specific representation $h_{A_i}^{r_k}$:
\begin{equation}
\label{equation6}
 h_{A_{i}\rightarrow B} = \frac{1}{K}{\sum_{k=1}^{K}\left(h_{A_i}^{r_k}\right)}.
\end{equation}
Note that $h_{A_{i}\rightarrow B}$ is recurrently updated with Eq.~\ref{equation3} and \ref{equation6}.
\end{enumerate}

\noindent%
Using Eq.~\ref{equation3} and \ref{equation6}, we obtain a sequence of representations $\langle h_{A_{1}\rightarrow B}, \ldots, h_{A_{N}\rightarrow B}\rangle$. 
We need to combine and transfer $\langle h_{A_{1}\rightarrow B}, \ldots, h_{A_{N}\rightarrow B}\rangle$ to domain $B$. 
We achieve this by employing another \ac{GRU} to recurrently encode $h_{A_{i}\rightarrow B}$ at each timestep to obtain $h_{(A\rightarrow B)_i}$, as shown in Eq.~\ref{equation7}: 
\begin{equation}
\begin{split}
\label{equation7}
h_{(A\rightarrow B)_i} = \operatorname{GRU}(h_{A_{i}\rightarrow B}, h_{(A\rightarrow B)_{i-1}}),
\end{split}
\end{equation}
where $h_{A_{i}\rightarrow B}$ is the representation filtered from domain $A$; $h_{(A\rightarrow B)_{i-1}}$ is the previous transformed representation at timestamp $i-1$. 
The initial state is also set to zero vectors, i.e., $h_{(A\rightarrow B)_0}=0$.
We set the cross-domain representation from domain $A$ to domain $B$ (i.e., $h_{(A\rightarrow B)}$) as the last timestamp representation $h_{(A\rightarrow B)_N}$, where $N$ is sequence length of domain $A$.


\begin{figure*}[t]
\centering
\includegraphics[width=0.65\textwidth]{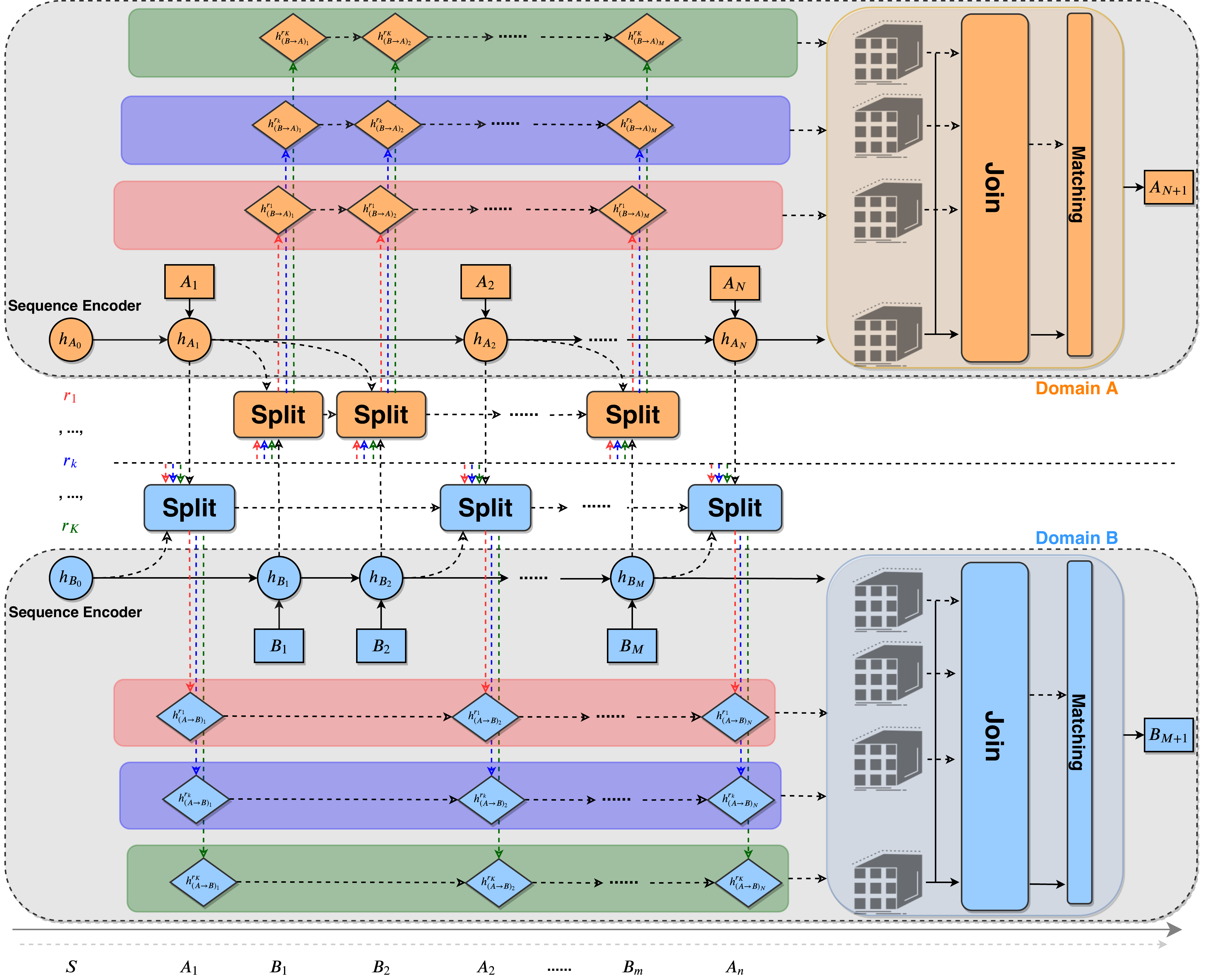}
\vspace*{0mm}
\caption{An overview of \ac{PSJNet}-\uppercase\expandafter{\romannumeral2}. 
As before, the orange and blue colors represent different domains.
Red, purple and green represent different user roles.
Section~\ref{section:psi-ii-walkthrough} contains a walkthrough of the model.}
\label{04}
\end{figure*}

\subsection{\ac{PSJNet}-\uppercase\expandafter{\romannumeral2}}
\label{section:psi-ii-walkthrough}

In this subsection, we describe \ac{PSJNet}-\uppercase\expandafter{\romannumeral2}, our second solution for \acs{SAC-SR}, in detail.
Unlike \ac{PSJNet}-\uppercase\expandafter{\romannumeral1}, \ac{PSJNet}-\uppercase\expandafter{\romannumeral2} is a ``Split-and-Join'' framework, which means that it first splits role-specific representations from the mixed user behavior at each timestamp.
Then the role-specific representations are transformed to another domain.
Finally, it joins the role-specific representations as cross-domain representations.
Figure~\ref{04} provides an overview of \ac{PSJNet}-\uppercase\expandafter{\romannumeral2}.
\ac{PSJNet}-\uppercase\expandafter{\romannumeral2} consists of four main components: 
\begin{enumerate*}[label=(\arabic*)]
\item a \textit{sequence encoder} (see Section~\ref{section:sequence-encoder}), 
\item a \textit{split unit}, 
\item a \textit{join unit}, and 
\item a \textit{hybrid recommendation decoder} (see Section~\ref{section:recommendation-decoder}). 
\end{enumerate*}
\ac{PSJNet}-\uppercase\expandafter{\romannumeral2} uses the same \textit{sequence encoder} and \textit{matching decoder} architectures as \ac{PSJNet}-\uppercase\expandafter{\romannumeral1}. 
Please refer to Section~\ref{section:sequence-encoder} and \ref{section:recommendation-decoder} for details of the \textit{sequence encoder} and the \textit{hybrid recommendation decoder}.
In this subsection, we focus on the core modules (i.e., the \textit{split unit} and \textit{join unit}) of \ac{PSJNet}-\uppercase\expandafter{\romannumeral2}.

\begin{figure*}
\centering
\includegraphics[width=0.65\textwidth]{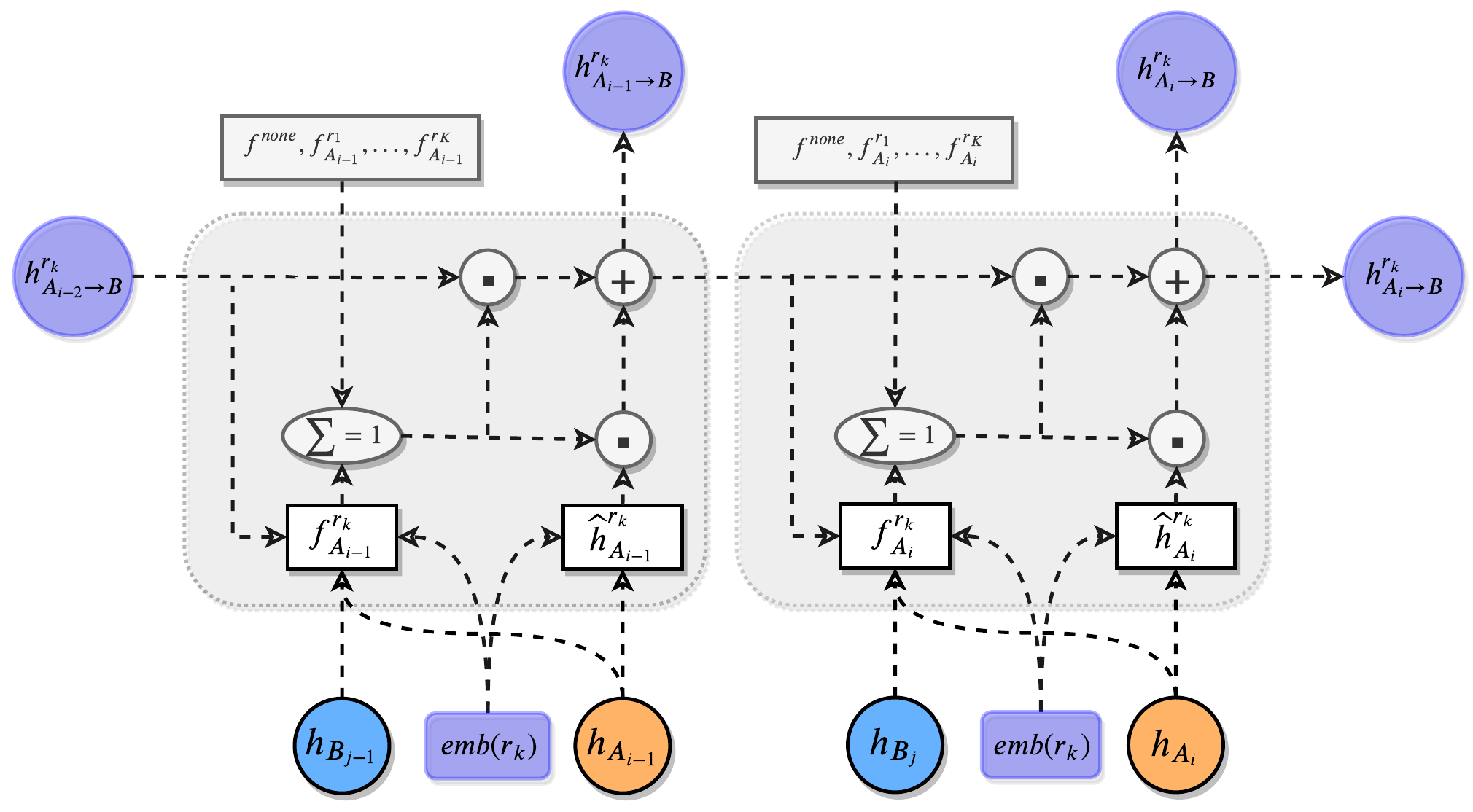}
\caption{The \textit{split unit} for $r_k$ illustrated while transforming information from domain $A$ to domain $B$.}
\label{05}
\end{figure*}

\subsubsection{Split unit}
The \textit{split unit} is shown in Figure~\ref{05}.
The differences with the \textit{split-by-join unit} of \ac{PSJNet}-\uppercase\expandafter{\romannumeral1} are marked in yellow.
As with \ac{PSJNet}-\uppercase\expandafter{\romannumeral1}, \ac{PSJNet}-\uppercase\expandafter{\romannumeral2} also assumes that there are $K$ latent roles under each account.
We split the specific representation for role $r_k$ at timestamp $i$ in domain $A$ with Eq.~\ref{equation8}:
\begin{equation}
\label{equation8}
h^{r_k}_{A_i\rightarrow B}= f_{A_i}^{r_k}\odot \widehat{h}_{A_i}^{r_k} + f_{A_i}^{none}\odot h^{r_k}_{A_{i-1}\rightarrow B},
\end{equation}
where $f_{A_i}^{none}$ is a special gate that handles the case when none of the information from $r_k$ at $i$ (i.e., $\widehat{h}_{A_i}^{r_k}$) is useful and we should inherit more information from previous time steps, see Eq.~\ref{equation9}:
\begin{equation}
\label{equation9}
f_{A_i}^{none}=\sigma \left(W_{f_A} \cdot h_{A_i} + W_{f_B} \cdot h_{B_j} + U_f \cdot h_{A_{i-1}\rightarrow B} + b_f\right).
\end{equation}
We add a normalization constraint to force the sum of $f_{A_i}^{r_k}$ and $f_{A_i}^{none}$ to $1$:
\begin{equation}
f_{A_i}^{none} + \sum_{k=1}^{K}{f_{A_i}^{r_k}} = 1.
\label{equation10}
\end{equation}
We use similar definitions of $f_{A_i}^{r_k}$ (Eq.~\ref{equation4}) and $\widehat{h}_{A_i}^{r_k}$ (Eq.~\ref{equation5}) as in \ac{PSJNet}-\uppercase\expandafter{\romannumeral1}, except that $h_{A_{i-1}\rightarrow B}$ is replaced with $h^{r_k}_{A_{i-1}\rightarrow B}$.
The differences from \textit{split-by-join unit} are two-fold.
First, $h^{r_k}_{A_i\rightarrow B}$ is not joined with respect to all roles.
Second, instead of learning independent gates for different roles, we require that all gate values from all roles (and $f_{A_i}^{none}$) are summed to $1$.

After Eq.~\ref{equation8}, we get a sequence of representations $\langle h^{r_k}_{A_{1}\rightarrow B}, \ldots, h^{r_k}_{A_{n}\rightarrow B}\rangle$ for each user role $r_k$.
We combine and transfer $\langle h^{r_k}_{A_{1}\rightarrow B}, \ldots, h^{r_k}_{A_{n}\rightarrow B}\rangle$ to domain $B$ by employing another \ac{GRU}, as shown in Eq.~\ref{equation11}: 
\begin{equation}
\label{equation11}
h^{r_k}_{(A\rightarrow B)_i} = \operatorname{GRU}(h^{r_k}_{A_{i}\rightarrow B}, h^{r_k}_{(A\rightarrow B)_{i-1}}),
\end{equation}
where $h^{r_k}_{A_{i}\rightarrow B}$ is the representation filtered from domain $A$ for role $r_k$.

\begin{figure*}
\centering
\includegraphics[width=0.65\textwidth]{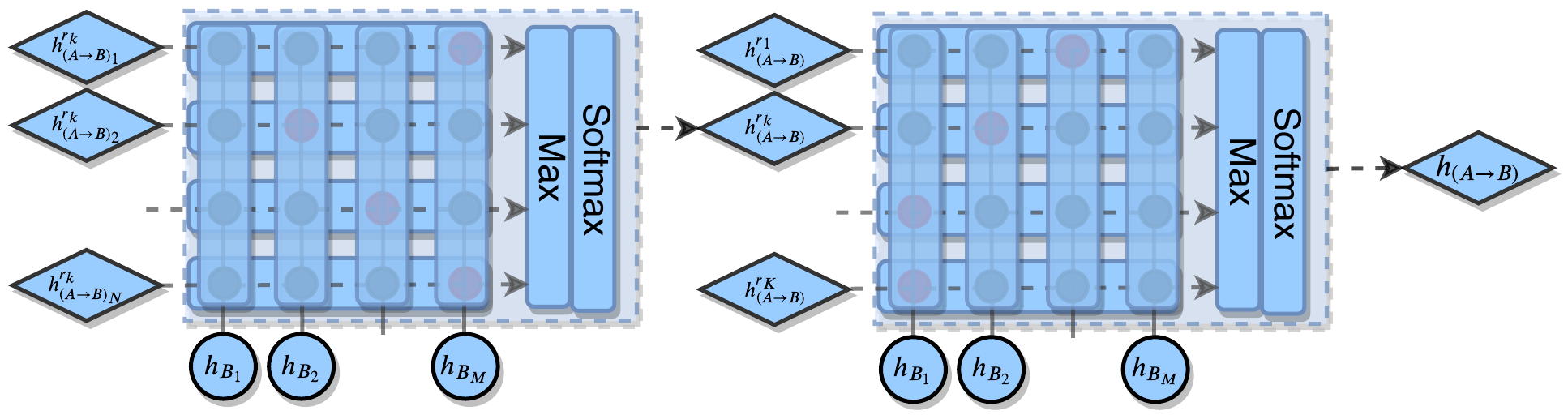}
\caption{The \textit{join unit} illustrated while transforming information from domain $A$ to domain $B$.}
\label{06}
\end{figure*}

\subsubsection{Join unit}
The \textit{join unit} is shown in Figure~\ref{06}.
After the \textit{split unit}, we obtain $K$ sequences of transformed representations $\langle h^{r_k}_{(A\rightarrow B)_1}, \ldots, h^{r_k}_{(A\rightarrow B)_{N}}\rangle$ from domain $A$ to domain $B$.
To join them, we first compute a similarity matrix $S^{\uppercase\expandafter{\romannumeral1}} \in \mathbb{R}^{M \times N}$ between the transformed representations and the in-domain representations $\langle h_{B_1}, \ldots, h_{B_M}\rangle$ from domain $B$.
Each similarity $S^{\uppercase\expandafter{\romannumeral1}}_{(i,j)}$ is computed with Eq.~\ref{equation12}:
\begin{equation}
\label{equation12}
S^{\uppercase\expandafter{\romannumeral1}}_{(i,j)} = {v_S}^{T} \cdot (W_i \cdot h^{r_k}_{(A\rightarrow B)_{i}} + W_j \cdot h_{B_j}),
\end{equation}
where ${v_S}^{T}$, $W_i$ and $W_j$ are parameters.

Then we pick the maximum similarity $S^{\uppercase\expandafter{\romannumeral1}}_i$ between each $h^{r_k}_{(A\rightarrow B)_{i}}$ and all $h_{B_j}$.
$S^{\uppercase\expandafter{\romannumeral1}}_i$ signifies that $h^{r_k}_{(A\rightarrow B)_{i}}$ is more representative for role $r_k$ in domain $B$ because it has the closest similarity to a representation $h_{B_j}$ in domain $B$:
\begin{equation}
S^{\uppercase\expandafter{\romannumeral1}}_i = \max_j{S^{\uppercase\expandafter{\romannumeral1}}_{(i,j)}}.
\end{equation}
We normalize $S^{\uppercase\expandafter{\romannumeral1}}_i$ with softmax afterwards.
Then we obtain representations for each role $r_k$ in Eq.~\ref{equation13}:
\begin{equation}
\label{equation13}
h^{r_k}_{(A\rightarrow B)}= \sum_{i=1}^{N} (S^{\uppercase\expandafter{\romannumeral1}}_i h^{r_k}_{(A\rightarrow B)_i}).
\end{equation}
Finally, we get the cross-domain representation $h_{(A\rightarrow B)}$ by joining $\langle h^{r_1}_{(A\rightarrow B)}, \ldots, h^{r_K}_{(A\rightarrow B)}\rangle$ again with similar operations as in \ref{equation12} and \ref{equation13}, but with a different similarity matrix $S^{\uppercase\expandafter{\romannumeral2}} \in \mathbb{R}^{M \times K}$.
Note that $S^{\uppercase\expandafter{\romannumeral2}}$ is computed between $\langle h^{r_1}_{(A\rightarrow B)}, \ldots, h^{r_K}_{(A\rightarrow B)}\rangle$ and $\langle h_{B_1}, \ldots, h_{B_M}\rangle$ this time.

There are two strengths of \ac{PSJNet}-\uppercase\expandafter{\romannumeral2} compared to \ac{PSJNet}-\uppercase\expandafter{\romannumeral1}. 
First, the normalization (see Eq.~\ref{equation10}) reduces the influence of some large gate values, thereby making the prediction more accurate. 
Second, the split-by-join unit of \ac{PSJNet}-\uppercase\expandafter{\romannumeral1} uses the output of the last time step of GRU as the cross-domain representation from domain $A$ to domain $B$. Information in the intermediate step is lost to some degree. 
However, in the join unit of \ac{PSJNet}-\uppercase\expandafter{\romannumeral2}, the cross-domain representation from A to B undergoes more fine-grained calculations.

\subsection{Hybrid recommendation decoder}
\label{section:recommendation-decoder}
The \textit{hybrid recommendation decoder} integrates hybrid information from both domains $A$ and $B$ to evaluate the recommendation probabilities of the candidate items.
Specifically, it first gets the hybrid representation by concatenating the representation $h_{B}$ from domain $B$ and the transformed representation $h_{(A\rightarrow B)}$ from domain $A$ to domain $B$. 
Then, it evaluates the recommendation probability for each candidate item by calculating the matching of between the hybrid representation and the item-embedding matrix followed by a softmax function, as shown in Eq.~\ref{equation15}:
\begin{equation}
\begin{split}
\label{equation15}
P(B_{j+1}|S) = \softmax\left(W_{I} \cdot \begin{bmatrix} h_{B}, h_{(A\rightarrow B)}\end{bmatrix}^\mathsf{T} +b_I\right),
\end{split}
\end{equation}
where $W_I$ is the embedding matrix of all items of domain $B$; $b_{I}$ is the bias term. 

\subsection{Objective function}
We employ the negative log-likelihood loss function to train \ac{PSJNet} in each domain as follows:
\begin{equation}
\begin{split}
& L_{A}(\theta)=-\frac{1}{|\mathbb{S}|}\sum_{S \in \mathbb{S}}\sum_{A_i \in S}\log P(A_{i+1}\mid S) \\
& L_{B}(\theta)=-\frac{1}{|\mathbb{S}|}\sum_{S \in \mathbb{S}}\sum_{B_j \in S}\log P(B_{j+1}\mid S),\\
\end{split}
\end{equation}
where $\theta$ are all the parameters of our model \ac{PSJNet} and $\mathbb{S}$ are the sequence instances in the training set.
In the case of joint learning, the final loss is a linear combination of both losses:
\begin{equation}
\begin{split}
& L(\theta)= L_{A}(\theta)+L_{B}(\theta).\\
\end{split}
\end{equation}
All parameters as well as the item embeddings are learned in an end-to-end back-propagation training way.


\section{Experimental Setup}

\subsection{Research questions}
We seek to answer the following research questions in our experiments: 
\begin{enumerate}[leftmargin=*, nosep, label=(RQ\arabic*)]
\item What is the performance of \ac{PSJNet}-\uppercase\expandafter{\romannumeral1} and \ac{PSJNet}-\uppercase\expandafter{\romannumeral2} on the \acs{SAC-SR} task? Do they outperform the state-of-the-art methods in terms of Recall and MRR? 
\item Which \ac{PSJNet} variant is more effective in the \acs{SAC-SR} task? \ac{PSJNet}-\uppercase\expandafter{\romannumeral1} or \ac{PSJNet}-\uppercase\expandafter{\romannumeral2}? What are the performances of different groups of methods, e.g., sequential and non-sequential recommendation methods? 
\item What is the performance of \ac{PSJNet}-\uppercase\expandafter{\romannumeral1} and \ac{PSJNet}-\uppercase\expandafter{\romannumeral2} on different domains and different datasets? Do they improve the performance of both domains and datasets? Are the improvements equivalent? 
\end{enumerate}

\subsection{Datasets}
\label{section:dataset}

We need datasets that exhibit both share-account and cross-domain characteristics to conduct experiments.
To demonstrate the effectiveness of the proposed \ac{PSJNet} model, we build and release two new datasets, HAmazon and HVIDEO, respectively.
We build the HAmazon dataset by simulating shared account characteristics using previously released Amazon datasets.
HVIDEO has previously been used in~\citep{ma-2019-pi-net} but we release it with this paper.
Details of the two datasets are as follows.

\begin{itemize}[leftmargin=*]
\item HAmazon: \citet{He:2016:UDM:2872427.2883037} have released an Amazon product dataset that contains product reviews (ratings, text, helpfulness votes) and metadata (descriptions, category information, price, brand, and image features) from Amazon; it includes 142.8 million reviews spanning the period May 1996--July 2014.
The data contains user behavior from multiple domains.
In this paper, we use data from two Amazon domains.
The M-domain contains \emph{movie} watching and rating behavior of Amazon users. 
The B-domain covers \emph{book} reading and rating behavior of Amazon users.
We collect user-id, item-id, rating, and timestamp from the data and conduct some preprocessing. 
We first order the items by time under each account.
Then, we merge records of the same item watched/read by the same user with an adjacent timestamp.
We only keep items whose frequency is larger than 5 in the M-domain and 10 in the B-domain.

To satisfy cross-domain characteristics, we first find users whose behavior can be found in both the Amazon movie and book domains and then only keep users who have more than 10 records.

To simulate shared account characteristics, we first split the data into 6 consecutive intervals, 1996--2000, 2001--2003, 2004--2006, 2007--2009, 2010--2012, and 2013--2015.
Then, we combine data from both domains by randomly merging 2, 3, or 4 users and their behavior in each interval as one shared account.
Because each sequence has a lot of user behavior recorded over a long period of time, we split the sequences from each account into several small sequences with each containing watching/reading records within a year.
We also require that each sequence contains at least 5 items from the M-domain and 2 items from the B-domain.
The length of each sequence is between 4 and 60 with an average length of 31.29.

For evaluation, we use the last watched/read item in each sequence for each domain as the ground truth respectively. 

We randomly select 75\% of all data as the training set, 15\% as the validation set, and the remaining 10\% as the test set.

The statistics of the final dataset are shown in Table~\ref{table1}.
Note that although HAmazon can be used for experiments, it is not a \ac{SAC-SR} dataset by nature.
There are two shortcomings.
First, the merged users do not naturally have the shared account characteristic.
Second, the two domains are quite different and are not well correlated in content, which means that the items in different domains have little chance to reflect similar interests.

\begin{table*}
\centering
\caption{Statistics of the datasets.}
\label{table1}
\begin{tabular}{lrclr}
\toprule
\multicolumn{2}{c}{HAmazon} && \multicolumn{2}{c}{HVIDEO}      \\ 
\cmidrule{1-2}
\cmidrule{4-5}
\emph{M-domain}         &     && \emph{V-domain}        &         \\ 
\#Items                 & 67,161  && \#Items                & 16,407  \\ 
\#Logs                  & 4,406,924  && \#Logs                 & 227,390 \\ 
\cmidrule{1-2}
\cmidrule{4-5}
\emph{B-domain}         &     && \emph{E-domain}        &         \\ 
\#Items                 & 126,547  && \#Items                & 3,380   \\ 
\#Logs                  & 4,287,240  && \#Logs                 & 177,758 \\ 
\cmidrule{1-2}
\cmidrule{4-5}
\#Overlapped-users      & 13,724  && \#Overlapped-users     & 13,714  \\ 
\#Sequences             & 289,160  && \#Sequences            & 134,349 \\ 
\#Training-sequences    & 204,477  && \#Training-sequences   & 102,182 \\ 
\#Validation-sequences  & 49,814  && \#Validation-sequences & 18,966  \\
\#Test-sequences \hspace*{2cm}        & 34,869  && \#Test-sequences \hspace*{2cm}       & 13,201  \\ \bottomrule
\end{tabular}
\end{table*}

\item HVIDEO: To facilitate future research for \ac{SAC-SR}, we also release another dataset, HVIDEO, which exhibits shared-account and cross-domain characteristics by nature.
HVIDEO is a smart TV dataset that contains watching logs of 260k users from October 1st 2016 to June 30th 2017. 
The logs are collected on two platforms (the V-domain and the E-domain) from a well-known smart TV service provider.
The V-domain contains family \emph{video} watching behavior including TV series, movies, cartoons, talent shows and other programs. 
The E-domain covers online \emph{educational} videos based on textbooks from elementary to high school, as well as instructional videos on sports, food, medical, etc.

On the two platforms, we gather user behavior, including which video is played, when a smart TV starts to play a video, and when it stops playing the video, and how long the video has been watched.
Compared with previous datasets, HVIDEO contains rich and natural family behavior data generated in a natural shared account and cross-domain context.
Therefore, it is very suitable for \acs{SAC-SR} research. 

We conduct some preprocessing on the dataset. 
We first filter out users who have less than 10 watching records and whose watching time is less than 300 seconds.
Then, we merge records of the same item watched by the same user with an adjacent time less than 10 minutes. 
After that, we combine data from different domains together in chronological order which is grouped by the same account. 
Because each account has a lot of logs recorded in a long time, we split the logs from each account into several small sequences with each containing 30 watching records.
And we require that the number of items in both domains must be greater than 5 in each sequence, which can ensure the sequences mix is high enough.

For evaluation, we use the last watched item in each sequence for each domain as the ground truth, respectively. 

We randomly select 75\% of all data as the training set, 15\% as the validation set, and the remaining 10\% as the test set.
The statistics of the final dataset are shown in Table~\ref{table1}.
\end{itemize}

\subsection{Baseline methods}
For our contrastive experiments, we consider baselines from four categories: traditional, sequential, shared account, and cross-domain recommendations. 

\subsubsection{Traditional recommendations.}
As traditional recommendation methods, we consider the following:

\begin{itemize}[leftmargin=*,nosep]
\item POP: This method ranks items in the training set based on their popularity, and always recommends the most popular items. 
It is a very simple baseline, but it can perform well in certain domains and is frequently used as a baseline because of its simplicity~\cite{no:33}. 
\item Item-KNN: The method computes a degree of item-to-item similarity that is defined as the number of co-occurrences of two items within sequences divided by the square root of the product of the number of sequences in which either item occurs. 
Items that are similar to the actual item will be recommended by this baseline. 
Regularization is included to avoid coincidental high similarities between rarely visited items~\cite{no:3}. 

\item BPR-MF: This model is a commonly used matrix factorization method.
This model cannot be applied directly to \acp{SR}, because new sequences do not have pre-computed feature vectors. 
Like~\cite{no:19}, we apply it for sequential recommendations by representing a new sequence with the average latent factors of items that appeared in this sequence, i.e., we average the similarities of the feature vectors between a recommendable item and the items of the session so far.
\end{itemize}

\subsubsection{Shared account recommendations.}
There are some studies that explore shared account recommendations by first achieving user identification \cite{no:49,no:54,no:51}.
However, they need extra information for user identification, e.g., some explicit ratings for movies or descriptions for some songs, even some textual descriptions for flight tickets, which is not available in our datasets.
Therefore, we select a method that works on the IP-TV recommendation task that is similar to ours.

\begin{itemize}[leftmargin=*,nosep]
\item VUI-KNN: This model encompasses an algorithm to decompose members in a composite account by mining different user interests over different time periods from logs~\cite{no:53}.
The method first divides a day into time periods, so the logs of each account fall into the corresponding time period; logs in each time period are assumed to be generated by a virtual user that is represented by a 3-dimensional $\lbrace account \times item \times time \rbrace $ vector. 
After that, cosine similarity is used to calculate similarity among virtual users, some of which are merged into a latent user. 
VUI deploys the User-KNN method to produce recommendations for these latent users.
\end{itemize}

\subsubsection{Cross-domain recommendations.}
For cross-domain recommendations, we choose two baseline methods.
\begin{itemize}[leftmargin=*,nosep]
\item NCF-MLP++: This model uses a deep learning-based process to learn the inner product of the traditional collaborative filtering by using a \ac{MLP} \cite{no:33}. 
We improve NCF-MLP by sharing the collaborative filtering in different domains. 
It is too time-consuming to rank all items with this method, because it needs to compute a score for each item one by one.
We randomly sample four negative instances for each positive instance in the training process, and sample 3,000 negatives for each predicted target item in the test process, thus simplifying the task for this method.
\item Conet: This is a neural transfer model across domains on the basis of a cross-stitch network \cite{no:34,misra2016cross}, where a neural collaborative filtering model~\cite{no:33} is employed to share information between domains.
\end{itemize}

\subsubsection{Sequential recommendations.}
Recently, a number of sequential recommendations methods have been proposed; \ac{RNN}-based neural methods have outperformed traditional \ac{MC}- or \ac{MDP}-based methods.
There are many \ac{RNN}-based methods.
In this paper, we consider two methods with somewhat similar architectures as \ac{PSJNet}.
\begin{itemize}[leftmargin=*,nosep]
\item GRU4REC: This model uses a \ac{GRU} to encode sequential information.
It uses a session-parallel mini-batch training process and employs ranking-based loss functions for learning the model \cite{no:19}.
\item HGRU4REC: \citet{no:1} propose this model based on \acp{RNN} which
can deal with two cases: 
\begin{enumerate*}[label=(\arabic*)]
\item user identifiers are present and propagate information from the previous sequence (user session) to the next, thus improving the recommendation accuracy, and 
\item there are no past sessions (i.e., no user identifiers). 
\end{enumerate*}
The model is based on a hierarchical \ac{RNN}, where the hidden state of a lower-level \ac{RNN} at the end of one sequence is passed as input to a higher-level \ac{RNN}, which is meant to predict a good initialization for the hidden state of the lower \ac{RNN} for the next sequence.
\end{itemize}

\subsection{Evaluation metrics}
Recommender systems can only recommend a limited number of items at a time. 
The item a user might pick should be amongst the first few in the ranked list \citep{no:1,no:58,no:59}. 
Commonly used metrics are MRR@20 and Recall@20 \citep{no:25,no:48,no:60}. 
In this paper, we also report MRR@5, Recall@5 and MRR@10, Recall@10. 

\begin{itemize}[leftmargin=*,nosep]
\item Recall: The primary evaluation metric is Recall, which measures the proportion of cases when the relevant item is amongst the top ranked items in all test cases. 
Recall does not consider the actual rank of the item as long as it is amongst the recommendation list.
This accords with certain real-world scenarios well where there is no highlighting of recommendations and the absolute order does not matter \cite{no:19}.

\item MRR: Another used metric is MRR (Mean Reciprocal Rank), which is the average of reciprocal ranks of the relevant items. 
And the reciprocal rank is set to zero if the ground truth item is not in the list of  recommendations. 
MRR takes the rank of the items into consideration, which is vital in settings where the order of recommendations matters.
We choose MRR instead of other ranking metrics, because there is only one ground truth item for each recommendation; no ratings or grade levels are available.
\end{itemize}

\noindent%
For significance testing we use a paired t-test with $p < 0.05$.

\subsection{Implementation details}
We set the item embedding size and GRU hidden state size to 90.
We use dropout \citep{srivastava2014dropout} with drop ratio $p$ = 0.8. 
These settings are chosen with grid search on the validation set.
For the latent user size $K$, we try different settings, an analysis of which can be found in Section~\ref{hyperparameter_analysis}.
We initialize the model parameters randomly using the Xavier method \citep{glorot2010understanding}. 
We take Adam as our optimizing algorithm. 
For the hyper-parameters of the Adam optimizer, we set the learning rate $\alpha = 0.001$. 
We also apply gradient clipping \citep{pascanu2013difficulty} with range $[-5, 5]$ during training. 
To speed up the training and converge quickly, we use mini-batch size 64. 
We test the model performance on the validation set for every epoch.
Both \ac{PSJNet}-\uppercase\expandafter{\romannumeral1} and \ac{PSJNet}-\uppercase\expandafter{\romannumeral2} are implemented in Tensorflow and trained on a GeForce GTX TitanX GPU.


\section{Experimental Results}
\label{result_sec}

\begin{table*}
\centering
\caption{Experimental results (\%) on the HAmazon dataset.}
\label{table2}
\begin{tabular}{lcccccccccccc}
\toprule
\multirow{3}{*}{\bf Methods} & \multicolumn{6}{c}{\bf M-domain recommendation} & \multicolumn{6}{c}{\bf B-domain recommendation}     \\
\cmidrule(r){2-7}\cmidrule{8-13}
& \multicolumn{3}{c}{MRR} & \multicolumn{3}{c}{Recall} & \multicolumn{3}{c}{MRR} & \multicolumn{3}{c}{Recall}   \\\cmidrule(r){2-4}\cmidrule(r){5-7}\cmidrule(r){8-10}\cmidrule{11-13}
& @5    & @10       & @20          & @5   & @10   & @20       & @5  & @10  & @20         & @5  & @10  & @20         \\
\midrule
POP &  \phantom{0}0.36    &  \phantom{0}0.44   &  \phantom{0}0.49  &  \phantom{0}0.73  &   \phantom{0}1.32  &  \phantom{0}2.02  &
\phantom{0}0.14 &   \phantom{0}0.19  & \phantom{0}0.22     & \phantom{0}0.42  & \phantom{0}0.78  &  \phantom{0}1.22     \\

Item-KNN  &    \phantom{0}1.28    &  \phantom{0}1.57  &  \phantom{0}1.86  &  \phantom{0}2.58  &  \phantom{0}4.83  & \phantom{0}9.00 & 
\phantom{0}3.23  & \phantom{0}3.94  &  \phantom{0}4.55  & \phantom{0}6.65 & 12.11  &   20.94     \\

BPR-MF    &   \phantom{0}2.90    & \phantom{0}3.00  &   \phantom{0}3.06  &    \phantom{0}3.90   &  \phantom{0}4.65   &  \phantom{0}5.50    &  \phantom{0}0.88   & \phantom{0}0.92   &  \phantom{0}0.96  &   \phantom{0}1.23   &   \phantom{0}1.50  & \phantom{0}2.15      \\

\midrule
VUI-KNN   &  --   &   --     &  --  & --   & -- & --  & -- &  --  & --   &  --  &   --   &  --  \\
\midrule
NCF-MLP++ &  13.68 &  13.96    &  14.21    &   18.44 &  20.58  &  24.31 & 
13.67   &  13.90 & 14.05   &   18.14  &   19.92    &   22.08  \\

Conet    & 14.64   &  14.90 & 15.12  &  19.25  & 21.25   & 24.46 & 
15.85  &  16.09 &  16.28   & 20.98   &  22.83  &  25.56   \\
\midrule
GRU4REC   &   82.01    &   82.08   & 82.11   &  83.10 &  83.61   &  84.06  &
 81.34 &   81.41  &  81.44  & 82.77  & 83.32   &  83.76  \\

HGRU4REC  &  83.07   & 83.12  & 83.14  & 84.24 &  84.65    &  84.91  &
  82.15 & 82.26  & 82.31  &  83.46 & 84.30  &  84.91    \\
\midrule
\ac{PSJNet}-\uppercase\expandafter{\romannumeral1}  & 83.91  & 83.94  &  83.95 & \textbf{84.91}  &  \textbf{85.13} & \textbf{85.33} & 
84.93  &  84.93 &  84.93  &  \textbf{85.33}  &  85.36 &  \textbf{85.38}  \\
\ac{PSJNet}-\uppercase\expandafter{\romannumeral2}  & \textbf{84.01}\rlap{$^\dagger$}  & \textbf{84.04}\rlap{$^\dagger$}  &  \textbf{84.05}\rlap{$^\dagger$} & 84.88  &  85.10 & 85.28 & 
\textbf{85.10}\rlap{$^\dagger$}  &  \textbf{85.10}\rlap{$^\dagger$} &  \textbf{85.11}\rlap{$^\dagger$}  &  85.32  &  \textbf{85.37} &  \textbf{85.38}  \\
\bottomrule
\end{tabular}%
\\[1.1ex]
\begin{minipage}{.8\textwidth}
\small{\textbf{Bold face} indicates the best result in terms of the corresponding metric.
Significant improvements over the best baseline results are marked with $^\dagger$ (t-test, $p < .05$). 
To ensure a fair comparison, we re-implemented GRUE4REC and HGRU4REC in Tensorflow, just like \ac{PSJNet}; the final results may be slightly different from the Theano version released by the authors. 
To obtain the results of NCF-MLP++ and Conet, we run the code released by \citet{no:34}.
VUI-KNN does not work on this dataset because it needs specific time in a day which is not available on the HAmazon dataset.
}
\end{minipage}
\end{table*}

\begin{table*}
\centering
\caption{Experimental results (\%) on the HVIDEO dataset.}
\label{table3}
\begin{tabular}{lcccccccccccc}
\toprule
\multirow{3}{*}{\bf Methods} & \multicolumn{6}{c}{\bf V-domain recommendation} & \multicolumn{6}{c}{\bf E-domain recommendation}     \\
\cmidrule(r){2-7}\cmidrule{8-13}
& \multicolumn{3}{c}{MRR} & \multicolumn{3}{c}{Recall} & \multicolumn{3}{c}{MRR} & \multicolumn{3}{c}{Recall}   \\\cmidrule(r){2-4}\cmidrule(r){5-7}\cmidrule(r){8-10}\cmidrule{11-13}
& @5    & @10       & @20          & @5   & @10   & @20       & @5  & @10  & @20         & @5  & @10  & @20         \\
\midrule
POP &  \phantom{0}2.66    &  \phantom{0}3.07   &  \phantom{0}3.27  &  \phantom{0}5.01  &   \phantom{0}7.77  &  10.49  &
\phantom{0}1.71 &   \phantom{0}1.96  & \phantom{0}2.24     & \phantom{0}2.21  & \phantom{0}3.61  &  \phantom{0}6.58      \\

Item-KNN  &    \phantom{0}4.43    &  \phantom{0}4.16  &  \phantom{0}2.93  &  10.48  &  16.49  & 23.93 & 
\phantom{0}2.11  & \phantom{0}2.39  &  \phantom{0}2.90  & \phantom{0}3.01 & \phantom{0}5.77  &   12.11     \\

BPR-MF    &   \phantom{0}1.21    & \phantom{0}1.31  &   \phantom{0}1.36  &    \phantom{0}1.88   &  \phantom{0}2.56   &  \phantom{0}3.38    &  \phantom{0}1.34   & \phantom{0}1.52   &  \phantom{0}1.64  &   \phantom{0}2.74   &   \phantom{0}4.05  & \phantom{0}5.83      \\

\midrule
VUI-KNN   &  \phantom{0}3.44   &   \phantom{0}3.53     &  \phantom{0}2.87  & 16.46   & 24.85 & 34.76  & 
\phantom{0}2.03 &  \phantom{0}2.51  & \phantom{0}3.48   &  \phantom{0}6.36  &   11.55   &  24.27  \\
\midrule
NCF-MLP++ &  16.25&  17.25    &  17.90    &   26.10 &  33.61  &  43.04 & 
\phantom{0}3.92   &  \phantom{0}4.57 & \phantom{0}5.14   &   \phantom{0}7.36  &   12.27    &   20.84  \\

Conet    & 21.25   &  22.61 & 23.28  &  32.94  & 43.07   & 52.72 & 
\phantom{0}5.01  &  \phantom{0}5.63 &  \phantom{0}6.21   & \phantom{0}9.26   &  14.07  &  22.71   \\
\midrule
GRU4REC   &   78.27    &   78.46   & 78.27   &  80.15 &  81.63   &  83.04  &
 12.27 &  13.00   &   13.70 &  16.24 &  21.89 &  32.16  \\

HGRU4REC  &  80.37   & 80.53  & 80.62  & 81.92 &  83.21    &  84.43  &
14.47   & 15.37  & 16.11  & 19.79  &  26.72 &  37.52    \\
\midrule
\ac{PSJNet}-\uppercase\expandafter{\romannumeral1}  & 80.51  & 80.80  &  80.95 & 83.22  &  85.34 & 87.48 & 
14.63  &  15.83 &  16.88  &  20.41  &  29.61 &  \textbf{45.19}  \\
\ac{PSJNet}-\uppercase\expandafter{\romannumeral2}  & \textbf{81.97}\rlap{$^\dagger$}  & \textbf{82.20}\rlap{$^\dagger$}  &  \textbf{82.32}\rlap{$^\dagger$} & \textbf{84.32}\rlap{$^\dagger$}  &  \textbf{86.11}\rlap{$^\dagger$} & \textbf{87.75}\rlap{$^\dagger$} & 
\textbf{16.63}\rlap{$^\dagger$}  &  \textbf{17.62}\rlap{$^\dagger$} &  \textbf{18.46}\rlap{$^\dagger$} &  \textbf{22.12}\rlap{$^\dagger$}  &  \textbf{29.64} &  42.20  \\
\bottomrule
\end{tabular}%
\\[1.1ex]
\begin{minipage}{0.8\textwidth}
The same conventions are used as in Table~\ref{table2}.
\end{minipage}
\end{table*}

To answer RQ1, RQ2 and RQ3, we report the results of \ac{PSJNet} compared with the baseline methods on the HAmazon and HVIDEO datasets, as shown in Table~\ref{table2} and \ref{table3}, respectively. 
From the tables, we can see that both \ac{PSJNet}-\uppercase\expandafter{\romannumeral1} and \ac{PSJNet}-\uppercase\expandafter{\romannumeral2} outperform all baselines in terms of MRR and Recall for all reported values. 
Below, we discuss several insights we obtain from Table~\ref{table2} and \ref{table3} so as to answer our research questions. 


\subsection{Overall performance on the \ac{SAC-SR} task (RQ1)}

Both two \ac{PSJNet} variants significantly outperform all baselines and achieve the best results on all metrics, including strong baselines, i.e., GRU4REC and HGRU4REC.
It is worth noting that although recent studies on \ac{SR} propose many neural network models, we choose GRU4REC and HGRU4REC because GRU4REC and HGRU4REC have very similar architectures as \ac{PSJNet}.
And to obtain a fair comparison, we re-implement them within the same TensorFlow framework as we use for \ac{PSJNet}.

Specifically, on the HVIDEO dataset, the largest increase of \ac{PSJNet}-\uppercase\expandafter{\romannumeral2} over GRU4REC is 4.04\% in terms of MRR@20, and 4.48\% in terms of Recall@10 on the V-domain.
On the E-domain, the increase is even larger with a 4.70\% increase of \ac{PSJNet}-\uppercase\expandafter{\romannumeral2} over GRU4REC in terms of MRR@20 and 13.03\% increase of \ac{PSJNet}-\uppercase\expandafter{\romannumeral1} over GRU4REC in terms of Recall@20.
And the increase over HGRU4REC on the V-domain is 1.69\% and 3.45\% (at most) in terms of MRR and Recall respectively. 
On the E-domain, the increase is 2.29\% and 7.67\% respectively.
We believe that those performance improvements are due to the fact that \ac{PSJNet} considers two important factors (shared-account and cross-domain) with its parallel modeling architecture and two main components for as part of its end-to-end recommendation model, namely the ``split'' and ``join''.
With these three modules, \ac{PSJNet} is able to model user interests more accurately by leveraging complementary information from both domains and improve recommendation performance in both domains. 
We will analyze the effects of the three modules in more depth in Section~\ref{AnalysisSAC}. 

\subsection{Comparing two versions of \ac{PSJNet} with different groups of methods (RQ2)}

Generally, \ac{PSJNet}-\uppercase\expandafter{\romannumeral2} outperforms \ac{PSJNet}-\uppercase\expandafter{\romannumeral1} on both datasets.
Specifically, \ac{PSJNet}-\uppercase\expandafter{\romannumeral2} outperforms \ac{PSJNet}-\uppercase\expandafter{\romannumeral1} in terms of most metrics on both domains on the HVIDEO dataset, especially for MRR@5 and Recall@5.
The performance is comparable on the HAmazon dataset.
But as we mentioned in \S\ref{section:dataset}, HAmazon is not a good dataset for \ac{SAC-SR} because the shared-account characteristic is simulated, and the two domains are quite different and not well correlated in content.
Since both \ac{PSJNet}-\uppercase\expandafter{\romannumeral1} and \ac{PSJNet}-\uppercase\expandafter{\romannumeral2} adopt the parallel modeling architecture, we can conclude that the superiority of \ac{PSJNet}-\uppercase\expandafter{\romannumeral2} over \ac{PSJNet}-\uppercase\expandafter{\romannumeral1} mainly comes from the separate modeling of ``split'' and ``join''.
We will show this in more depth in Section~\ref{AnalysisSAC}. 

We can also see that \ac{RNN}-based methods (e.g., GRU4REC, HGRU4REC, and our \ac{PSJNet}) perform much better than traditional methods, which demonstrates that \ac{RNN}-based methods are good at dealing with sequential information. 
They are able to learn better dense representations of the data through nonlinear modeling, which we assume is able to provide a higher level of abstraction. 
The shared account and cross-domain baselines (e.g., VUI-KNN, NCF-MLP++ and Conet) perform much worse than \ac{PSJNet}.
They also perform substantially worse than HGRU4REC.
This is because these shared account and cross-domain baselines ignore sequential information; VUI-KNN only considers the length of watching time, and NCF-MLP++ and Conet do not use any time information.
Another reason is that NCF-MLP++ and Conet just map each overlapped account in both domains to the same latent space to calculate the user-item similarity. 
However, the existence of shared accounts makes it difficult to find the same latent space for different latent user roles under one account.
Besides, VUI-KNN is not a deep learning method and it simply distinguishes user roles based on the fixed divided time periods in a day, which means it assumes there is only one family member in each time period.
However, in the smart TV scenario, many people usually watch programs together \cite{no:53}.
This situation cannot be captured very well by VUI-KNN.  
And it requires the specific time of the user behaviors in a day in order to distinguish different user roles.
That is why we cannot use it to obtain results on the HAmazon dataset because there is no such information.
In contrast, \ac{PSJNet} can extract components of each hidden user role according to their viewing records in another domain with the ``split'' module.
The results of BPR-MF are lower than of POP, which indicates that most items users watched are very popular, which is in line with the phenomenon that people like to pursue popular items in the video and book recommendation scenarios.

\subsection{Contrasting the performance on different domains and different datasets (RQ3)}

The Recall values of \ac{PSJNet} on the HAmazon dataset are comparable on the two domains while the Recall values on the V-domain are higher than those on the E-domain on the HVIDEO dataset. 
This is also true for the other methods.
We believe that this is because of data balance issues.
On the HAmazon dataset, the data is generally balanced on two domains.
Most accounts have an equal amount of data on both domains.
This means that the models can learn pretty well with data from just one domain.
Cross-domain information is not that important:
the increase of \ac{PSJNet} on the HAmazon dataset is relatively small.
However, the situation is different on the HVIDEO dataset.
Most accounts have much more data on the V-domain due to its content diversity; because of this, models can better learn user's viewing characteristics on the V-domain.
Therefore, on the HAmazon dataset, the space for improvement on both domains is smaller than on the HVIDEO dataset.

Additionally, comparing \ac{PSJNet} with the best baseline, HGRU4REC, we find that the largest increase on the E-domain is larger than on the V-domain.
The largest increase in MRR is 1.69\% on the V-domain and 2.29\% on the E-domain.
And for the Recall values, the largest increase is 3.45\% on the V-domain, and 7.67\% on the E-domain. 
This shows that the space for potential improvements on the V-domain is smaller than on the E-domain after considering shared account and cross-domain information.

Also, the increases in MRR and Recall are different on two datasets.
On the HAmazon dataset, there is no significant difference for both MRR and Recall from @5 to @20.
This means that \ac{PSJNet} can already predict the ground truth item within the top-5 for most cases.
This is not true on the HVIDEO dataset, especially on the E-domain.
Specifically, the largest increase is 2.25\% for MRR from the top-5 to the top-20, and 24.78\% for Recall.

\section{Analysis}

\begin{table*}
\centering
\caption{Analysis of the contribution of the parallel modeling, split unit and join unit on the HAmazon dataset.}
\label{table4}
\begin{tabular}{lcccccccccccc}
\toprule
\multirow{3}{*}{\bf Methods} & 
\multicolumn{6}{c}{\bf M-domain recommendation} & \multicolumn{6}{c}{\bf B-domain recommendation}  \\ 
\cmidrule(r){2-7}\cmidrule{8-13} 
& \multicolumn{3}{c}{MRR} & \multicolumn{3}{c}{Recall} & \multicolumn{3}{c}{MRR} & \multicolumn{3}{c}{Recall} \\ 
\cmidrule(r){2-4}\cmidrule(r){5-7}\cmidrule(r){8-10}\cmidrule{11-13} 
& @5     & @10    & @20    & @5      & @10      & @20    & @5     & @10    & @20    & @5      & @10      & @20    \\ \midrule
\ac{PSJNet} (-PSJ) & 77.26   & 77.44  &  77.51  & 82.22   & 83.52  & 84.39
&  81.69      &  81.72   & 81.73    &  85.03    & 85.27  & 85.34  \\

\ac{PSJNet}-\uppercase\expandafter{\romannumeral1} (-SJ) &   83.30    &   83.32    &     83.33   &   84.03    &   84.20       &  84.31    
&    84.04    &    84.04    &   84.04  &     85.31   &  85.35     &   \textbf{85.38}    \\ 
\ac{PSJNet}-\uppercase\expandafter{\romannumeral2} (-S) &   83.55    &   83.59    &     83.60   &   84.61    &   84.90       &  85.14    
&    84.87    &    84.88    &   84.88  &     85.26   &  85.31     &   85.35    \\ 
\ac{PSJNet}-\uppercase\expandafter{\romannumeral2} (-J) &   82.28    &   82.35    &     82.38   &   84.02   &   84.52       &  84.92 
&    83.42   &    83.45    &   83.46  &     84.79   &  84.96     &   85.08    \\ 
\ac{PSJNet}-\uppercase\expandafter{\romannumeral1}  & 83.91  & 83.94  &  83.95 & \textbf{84.91}  &  \textbf{85.13} & \textbf{85.33} & 
84.93  &  84.93 &  84.93  &  \textbf{85.33}  &  85.36 &  \textbf{85.38}  \\
\ac{PSJNet}-\uppercase\expandafter{\romannumeral2}  & \textbf{84.01}  & \textbf{84.04}  &  \textbf{84.05} & 84.88  &  85.10 & 85.28 & 
\textbf{85.10}  &  \textbf{85.10} &  \textbf{85.11}  &  85.32  &  \textbf{85.37} &  \textbf{85.38}  \\
\bottomrule
\end{tabular}%
\\[1.1ex]
\begin{minipage}{0.8\textwidth}
\small{
\ac{PSJNet} (-PSJ) is \ac{PSJNet} without parallel modeling, i.e., no cross-domain representations are used for recommendations.
Without parallel modeling, both \ac{PSJNet}-\uppercase\expandafter{\romannumeral1} and \ac{PSJNet}-\uppercase\expandafter{\romannumeral2} become the same \ac{PSJNet} (-PSJ).
\ac{PSJNet}-\uppercase\expandafter{\romannumeral1} (-SJ) is \ac{PSJNet}-\uppercase\expandafter{\romannumeral1} without ``split-by-join'' unit.
Because ``split-by-join'' is an indivisible unit, there is no \ac{PSJNet}-\uppercase\expandafter{\romannumeral1} (-S) or \ac{PSJNet}-\uppercase\expandafter{\romannumeral1} (-J).
\ac{PSJNet}-\uppercase\expandafter{\romannumeral2} (-S) is \ac{PSJNet}-\uppercase\expandafter{\romannumeral2} without the ``split'' unit and \ac{PSJNet}-\uppercase\expandafter{\romannumeral2} (-J) is \ac{PSJNet}-\uppercase\expandafter{\romannumeral2} without the ``join'' unit.
}
\end{minipage}
\end{table*}

\begin{table*}
\centering
\caption{Analysis of the contribution of the parallel modeling, split unit and join unit on the HVIDEO dataset.}
\label{table5}
\begin{tabular}{lcccccccccccc}
\toprule
\multirow{3}{*}{\bf Methods} & 
\multicolumn{6}{c}{\bf V-domain recommendation} & \multicolumn{6}{c}{\bf E-domain recommendation}  \\ 
\cmidrule(r){2-7}\cmidrule{8-13} 
& \multicolumn{3}{c}{MRR} & \multicolumn{3}{c}{Recall} & \multicolumn{3}{c}{MRR} & \multicolumn{3}{c}{Recall} \\ 
\cmidrule(r){2-4}\cmidrule(r){5-7}\cmidrule(r){8-10}\cmidrule{11-13} 
& @5     & @10    & @20    & @5      & @10      & @20    & @5     & @10    & @20    & @5      & @10      & @20    \\ \midrule
\ac{PSJNet} (-PSJ) & 78.02   & 78.17  &  78.28  & 80.13   & 81.34  & 82.93
&  12.69      &  13.43   & 14.05    &  16.54    & 22.29  & 31.45  \\

\ac{PSJNet}-\uppercase\expandafter{\romannumeral1} (-SJ) &   78.59    &   78.85    &     78.97   &   81.71    &   83.58       &  85.33    
&    16.35    &    17.04    &   17.59  &     20.97   &  26.29     &   34.44    \\ 
\ac{PSJNet}-\uppercase\expandafter{\romannumeral2} (-S) &   81.61    &   81.85    &     81.99   &   83.93    &   85.73       &  87.71    
&    15.94    &    17.01    &   17.84  &     20.96   &  29.18     &   41.38    \\ 
\ac{PSJNet}-\uppercase\expandafter{\romannumeral2} (-J) &   81.76    &   81.98    &     82.12   &   84.20   &   85.80       &  \textbf{87.77} 
&    16.43   &    17.48    &   \textbf{18.46}  &     21.92   &  \textbf{29.96}     &   44.30    \\ 
\ac{PSJNet}-\uppercase\expandafter{\romannumeral1}  & 80.51  & 80.80  &  80.95 & 83.22  &  85.34 & 87.48 & 
14.63  &  15.83 &  16.88  &  20.41  &  29.61 &  \textbf{45.19}  \\
\ac{PSJNet}-\uppercase\expandafter{\romannumeral2}  & \textbf{81.97}  & \textbf{82.20}  &  \textbf{82.32} & \textbf{84.32}  &  \textbf{86.11} & 87.75 & 
\textbf{16.63}  &  \textbf{17.62} &  \textbf{18.46} &  \textbf{22.12}  &  29.64 &  42.20  \\
\bottomrule
\end{tabular}%
\\[1.1ex]
\begin{minipage}{.8\textwidth}
The same settings are applied as in Table~\ref{table4}.
\end{minipage}
\end{table*}

\subsection{Ablation study}
\label{AnalysisSAC}

In this subsection, we report on an ablation study to verify how well the parallel modeling schema, with the ``split'' and ``join'' units, improves the recommendation performance.
The results are shown in Table~\ref{table4} and~\ref{table5}.
\ac{PSJNet} (-PSJ) is the \ac{PSJNet}-\uppercase\expandafter{\romannumeral1} or \ac{PSJNet}-\uppercase\expandafter{\romannumeral2} without all the three parts, which degenerates into GRU4REC except that \ac{PSJNet} (-PSJ) is jointly trained on two domains.
\ac{PSJNet}-\uppercase\expandafter{\romannumeral1} (-SJ) is \ac{PSJNet}-\uppercase\expandafter{\romannumeral1} without ``split-by-join'' unit.
\ac{PSJNet}-\uppercase\expandafter{\romannumeral2} (-S) is \ac{PSJNet}-\uppercase\expandafter{\romannumeral2} without the ``split'' unit and \ac{PSJNet}-\uppercase\expandafter{\romannumeral2} (-J) is \ac{PSJNet}-\uppercase\expandafter{\romannumeral2} without the ``join'' unit (i.e., replacing the ``join'' unit by summing up the outputs from the ``split'' unit).
We can draw the following conclusions from the results.

First, almost all the best results are almost all from \ac{PSJNet}-\uppercase\expandafter{\romannumeral1} and \ac{PSJNet}-\uppercase\expandafter{\romannumeral2}, which demonstrates the effectiveness of combining all three parts.
The three parts bring around 7\% (MRR) and 1\%--3\% (Recall) improvements on the M-domain of HAmazon, and around 4\% (MRR) and 4\%--10\% (Recall) on both domains of HVIDEO.
Additionally, we see that \ac{PSJNet} (-PSJ) gets the lowest performance amongst  these methods, while it still outperforms most of the baselines listed in Table~\ref{table2} and \ref{table3}.
In summary, then, combining information from an auxiliary domain is useful. 
The MRR improvements are larger on HAmazon while the Recall improvements are larger on HVIDEO.
This is due to the different characteristics of different domains.
Take the two domains in HVIDEO for example.
Almost all members have viewing records in the V-domain. 
However, items on the E-domain are mostly educational programs, so children take up a large proportion, and their educational interests are relatively fixed. 
As a result, the information extracted from the V-domain mostly belongs to children, which is less helpful because we already have enough data on the E-domain to learn such features in most cases. 

Second, generally parallel modeling brings the most improvements followed by the ``split'' and ``join'' units.
Specifically, \ac{PSJNet}-\uppercase\expandafter{\romannumeral1} achieves around 5\% (MRR) and 2\% (Recall) improvements on the M-domain of HAmazon with the parallel modeling while  further improvements with the ``split-by-join'' unit are just around 0.6\% (MRR) and 1\% (Recall).
Similar results can be found on the B-domain of HAmazon and E-domain of HVIDEO.
We believe this is because the model is already able to leverage information from both domains to achieve recommendations with the parallel modeling schema.
It is further improved by taking other factors, e.g., shared-account characteristics, into account in order to better leverage the cross-domain information.
This is why the ``split'' and ``Joint'' units are able to further improve the results over the parallel modeling schema.
An exception is that the ``split'' and ``join'' units achieve more improvements than the parallel modeling on the V-domain of HVIDEO for \ac{PSJNet}-\uppercase\expandafter{\romannumeral1}.
We think the reason is that \ac{PSJNet}-\uppercase\expandafter{\romannumeral1} (-SJ) cannot effectively use the cross-domain information without the ``split-by-join'' unit, while \ac{PSJNet}-\uppercase\expandafter{\romannumeral2} (-S) is better because the function of ``split'' unit is replaced by the ``join'' unit to some extent.
The same is true for  \ac{PSJNet}-\uppercase\expandafter{\romannumeral2} (-J).
This could be verified by the fact that both \ac{PSJNet}-\uppercase\expandafter{\romannumeral1} and \ac{PSJNet}-\uppercase\expandafter{\romannumeral2} get big improvements with both units than with neither, but the improvements are smaller than with one unit for \ac{PSJNet}-\uppercase\expandafter{\romannumeral2}.

Third, the ``split'' unit is generally more effective than the ``join'' unit for \ac{PSJNet}-\uppercase\expandafter{\romannumeral2} as we find that the gap between \ac{PSJNet}-\uppercase\expandafter{\romannumeral2} and \ac{PSJNet}-\uppercase\expandafter{\romannumeral2} (-J) is smaller than between \ac{PSJNet}-\uppercase\expandafter{\romannumeral2} and \ac{PSJNet}-\uppercase\expandafter{\romannumeral2} (-S).
On the one hand, this shows that the ``split'' unit plays a more important role which addresses the challenge raised by shared accounts, i.e., filtering out information of some user roles that might be useful for another domain from the mixed user behaviors.
On the other hand, the results also show that the current ``join'' unit is not effective enough as directly summing up the outputs from the ``split'' unit also achieves competitive performance, and/or the improvement space of the ``join'' unit is limited.

\subsection{Influence of the hyperparameter $K$}
\label{hyperparameter_analysis}

\begin{table*}
\centering
\caption{Analysis of the hyperparameter $K$ on the HAmazon dataset.}
\label{table6}
\begin{tabular}{ccccccccccccc}
\toprule
\multirow{3}{*}{\bf $K$ values}   & \multicolumn{6}{c}{\bf M-domain recommendation}             & \multicolumn{6}{c}{\bf B-domain recommendation}             \\ 
\cmidrule(r){2-7} \cmidrule{8-13} 
                           & \multicolumn{3}{c}{MRR} & \multicolumn{3}{c}{Recall} & \multicolumn{3}{c}{MRR} & \multicolumn{3}{c}{Recall} \\ 
\cmidrule(r){2-4} \cmidrule(r){5-7} \cmidrule{8-10} \cmidrule{11-13} 
                           & @5     & @10    & @20    & @5      & @10      & @20    & @5     & @10    & @20    & @5      & @10      & @20    \\ \midrule
\multicolumn{13}{c}{\bf \ac{PSJNet}-\uppercase\expandafter{\romannumeral1}} \\
1 &    82.45    &   82.52     &   82.54     &     84.23    &     84.69     &   85.07     & 
    84.72 &   84.73     &   84.73     &    85.29     &    85.35      &    85.38    \\ 
    
2 &   83.35     &    83.40    &     83.41   &     84.66    &    85.02      &   85.18     & 
    84.74 &   84.75     &   84.75     &    85.30     &    85.25      &    85.37    \\ 
    
3&    83.65    &  83.68      &    83.70    &   84.81      &    85.08      &    85.30    & 
    84.89 &   84.89     &    84.89    &    85.32     &     85.35     &   85.38     \\ 
    
4&     \textbf{83.91}   &    \textbf{83.94}    &    \textbf{83.95}    &    \textbf{84.91}     &    \textbf{85.13}      &  \textbf{85.33}      & 
   84.93  &   84.93     &   84.93     &    \textbf{85.33}     &    \textbf{85.40}      &   85.38     \\

5&    83.73   &   83.76    &   83.78   &   84.82      &    85.08      &   85.32     &  
    \textbf{84.94}  &    \textbf{84.94}    &   \textbf{84.94}     &    \textbf{85.33}     &    85.38      &     \textbf{85.39}   \\\midrule
\multicolumn{13}{c}{\bf \ac{PSJNet}-\uppercase\expandafter{\romannumeral2}} \\
1 &   84.33     &    \textbf{84.36}    &     \textbf{84.37}   &     \textbf{85.01}    &    \textbf{85.19}      &   \textbf{85.32}     & 
    85.09 &   85.10     &   85.10     &    85.32     &    85.36      &    85.39    \\ 
    
2 &   \textbf{84.08}     &    84.12    &     84.13   &     84.92    &    85.15      &   85.30     & 
    85.13 &   85.13     &   85.13     &    \textbf{85.33}     &    85.36      &    \textbf{85.40}    \\ 
    
3&    84.03    &  84.06      &    84.07    &   84.92      &    85.12      &    85.29    & 
    \textbf{85.16} &   \textbf{85.16}     &    \textbf{85.16}    &    \textbf{85.33}     &     85.35     &   85.37     \\ 
    
4&     84.01   &    84.04    &    84.05    &    84.88     &    85.10      &  85.28      & 
    85.10  &   85.10     &   85.11     &    85.32     &    \textbf{85.37}      &   85.38     \\

5&    82.34   &   82.42    &   82.44   &   84.06      &    84.63      &   84.99     &  
    84.67  &    84.68    &   84.69     &    85.23     &    85.30      &     85.37   \\
\bottomrule
\end{tabular}%
\end{table*}

\begin{table*}
\centering
\caption{Analysis of the hyperparameter $K$ on the HVIDEO dataset.}
\label{table7}
\begin{tabular}{ccccccccccccc}
\toprule
\multirow{3}{*}{\bf $K$ values}   & \multicolumn{6}{c}{\bf V-domain recommendation}             & \multicolumn{6}{c}{\bf E-domain recommendation}             \\ 
\cmidrule(r){2-7} \cmidrule{8-13} 
                           & \multicolumn{3}{c}{MRR} & \multicolumn{3}{c}{Recall} & \multicolumn{3}{c}{MRR} & \multicolumn{3}{c}{Recall} \\ 
\cmidrule(r){2-4} \cmidrule(r){5-7} \cmidrule{8-10} \cmidrule{11-13} 
                           & @5     & @10    & @20    & @5      & @10      & @20    & @5     & @10    & @20    & @5      & @10      & @20    \\ \midrule
\multicolumn{13}{c}{\bf \ac{PSJNet}-\uppercase\expandafter{\romannumeral1}} \\
1 &    80.19    &   80.50     &   80.66     &     82.85    &     85.15     &   87.40     & 
    13.92 &   15.06     &   16.10     &    19.76     &    28.74      &    43.98    \\ 
    
2 &   80.48     &    80.75    &     80.91   &     83.08    &    85.06      &   87.31     & 
    14.29 &   15.47     &   16.54     &    19.83     &    28.96      &    44.77    \\ 
    
3&    80.53    &  80.79      &    80.93    &   \textbf{83.34}      &    85.31      &    87.31    & 
    14.45 &   15.54     &    16.64    &    20.23     &     28.61     &   44.64     \\ 
    
4&     80.51   &    80.80    &    80.95    &    83.22     &    \textbf{85.34}      &  \textbf{87.48}      & 
    \textbf{14.63}  &   \textbf{15.83}     &   \textbf{16.88}     &    20.41     &    \textbf{29.61}      &   \textbf{45.19}     \\

5&    \textbf{80.60}   &   \textbf{80.86}    &   \textbf{81.02}   &   83.25      &    85.19      &   87.47     &  
    14.59  &    15.71    &   16.75     &    \textbf{20.42}    &    28.97      &     44.38   \\\midrule
\multicolumn{13}{c}{\bf \ac{PSJNet}-\uppercase\expandafter{\romannumeral2}} \\
1 &    81.93    &   82.18     &   82.32     &     \textbf{84.33}    &     \textbf{86.17}     &   \textbf{88.21}     & 
    16.17 &   17.18     &   18.13     &    21.42     &    29.23      &    \textbf{43.29}    \\ 
    
2 &   81.80     &    82.04    &     82.17   &     84.26    &    86.05      &   87.90     & 
    16.62 &   17.67     &   18.55     &    21.60     &    29.60      &    42.63    \\ 
    
3&    81.86    &  82.08      &    82.20    &   84.14      &    85.80      &    87.53    & 
    \textbf{16.90} &   \textbf{17.94}     &    \textbf{18.77}    &    \textbf{22.42}     &     \textbf{30.36}     &   42.51     \\ 
    
4&     \textbf{81.97}   &    \textbf{82.20}    &    \textbf{82.32}    &    84.32     &    86.11      &  87.75      & 
    16.63  &   17.62     &   18.46     &    22.12     &    29.64      &   42.20     \\

5&    81.78   &   82.02    &   82.14   &   83.99      &    85.67      &   87.68     &  
    16.78  &    17.84    &   18.66     &    22.01     &    30.07      &     42.13   \\
\bottomrule
\end{tabular}%
\end{table*}

Both \ac{PSJNet}-\uppercase\expandafter{\romannumeral1} and \ac{PSJNet}-\uppercase\expandafter{\romannumeral2} introduce a hyperparameter $K$ in the ``split'' unit which corresponds to the number of latent user roles.
We carry out experiments to study how setting $K$ affects the recommendation performance of \ac{PSJNet} on both datasets, and whether the best $K$ is the same under all situations and accords with reality.
Taking into account common sizes of families, we consider $K=1,\ldots, 5$, and compare different values of $K$ while keeping other settings unchanged.
The results are shown in Table~\ref{table6} and~\ref{table7}.

First, we see that the best values in terms of MRR and Recall are achieved when $K=3, 4$, $K=4$ for \ac{PSJNet}-\uppercase\expandafter{\romannumeral1} and $K=3$ for \ac{PSJNet}-\uppercase\expandafter{\romannumeral2} especially.
This is consistent with the size of modern families on HVIDEO and the simulation settings on HAmazon.
For \ac{PSJNet}-\uppercase\expandafter{\romannumeral1}, the lowest MRR and Recall values are achieved when $K=1$.
But for \ac{PSJNet}-\uppercase\expandafter{\romannumeral2}, the gap between the best and worst performances is much smaller, which indicates that \ac{PSJNet}-\uppercase\expandafter{\romannumeral2} is less sensitive to $K$ than \ac{PSJNet}-\uppercase\expandafter{\romannumeral1}.

Seond, both \ac{PSJNet}-\uppercase\expandafter{\romannumeral1} and \ac{PSJNet}-\uppercase\expandafter{\romannumeral2} show mostly consistent trends and conclusions on both datasets, i.e., the best $K$ values are basically the same.
On the one hand, this demonstrates the performance stability of both \ac{PSJNet}-\uppercase\expandafter{\romannumeral1} and \ac{PSJNet}-\uppercase\expandafter{\romannumeral2}.
On the other hand, this is also a clue that both models identify $K$ as the potential user roles under each account, which verifies our assumption.

Third, although $K$ can affect the recommendation performance, the influence is limited.
As we can see that the largest gaps between the best and worst performances are 1.94\% (MRR) and 0.56\% (Recall) on HAmazon, 0.78\% (MRR) and 1.21\% (Recall) on HVIDEO.
This is because even if $K=1,2$, our models still consider the information of all members except that some members are modeled as a single latent user role.

\section{Conclusion and Future Work}
We have studied the task of \acf{SAC-SR} and proposed an extension to our previous work~\cite{ma-2019-pi-net}.
We have generalized over the previous proposal ($\pi$-Net) with a more general framework that allows us to come up with a better performing model.
Under this framework, we have reformulated $\pi$-Net as \ac{PSJNet}-\uppercase\expandafter{\romannumeral1} and proposed a new instantiation, \ac{PSJNet}-\uppercase\expandafter{\romannumeral2}, with different split-join schemes.
Experimental results demonstrate that \ac{PSJNet} outperforms state-of-the-art methods in terms of MRR and Recall.
We have also conducted extensive analysis experiments to show the effectiveness of the two \ac{PSJNet} variants.

A limitation of \ac{PSJNet} is that it works better only when we have shared information in two domains that are complementary to each other.
Otherwise, \ac{PSJNet} only achieves comparable performance with state-of-the-art methods for shared account and/or cross-domain recommendations.

As to future work, \ac{PSJNet} can be advanced in several directions.
First, we assume the same number of latent user roles under each account in this study.
This can be further improved by automatically detecting the number of user roles, e.g., adaptively setting the number of family members in smart TV scenarios.
Second, we have focused on the architecture of \ac{PSJNet} and have not explored alternative choices for some of its main ingredients (e.g., encoders, decoders and loss functions).
It would be interesting to see whether alternative choices will further improve the performance of \ac{PSJNet}.
Third, side information (e.g., movie categories, attributes or labels, etc.) has been proven effective in improving recommendation performance in traditional recommendation \cite{Forsati:2014:MFE:2684820.2641564, Vasile:2016:MPE:2959100.2959160} and \ac{SR} \cite{Chen:2018:SRU:3159652.3159668}.
We hope to explore how to better incorporate side information into \ac{PSJNet} for \ac{SAC-SR}.
Fourth, explainability is seen as important challenge for deep learning at present.
Explainability is not the focus of this work,  it is interesting to see how effective explanations can be produced for different stakeholders in the complex domain of \acf{SAC-SR}~\citep{lucic-2021-multistakeholder}.


\section*{Acknowledgements}
This work is supported by the Hybrid Intelligence Center, 
a 10-year programme
funded by the Dutch Ministry of Education, Culture and Science through 
the Netherlands
Organisation for Scientific Research, 
\url{https://hybrid-intelligence-centre.nl}, the Natural Science Foundation of China (61672324, 61672322, 61972234, 61902219), the Natural Science Foundation of Shandong province (2016ZRE27468), the Tencent AI Lab
Rhino-Bird Focused Research Program (JR201932), and the Fundamental Research Funds of Shandong University. All content represents the opinion of the authors, which is not necessarily shared or endorsed by their respective employers and/or sponsors.

\section*{Code and data}
The code used to run the experiments in this paper is available at \url{https://bitbucket.org/Catherine_Ma/sequentialrec/src/master/tois-PsiNet/code/}.
The datasets released in this paper are shared at \url{https://bitbucket.org/Catherine_Ma/sequentialrec/src/master/tois-PsiNet/datasets/}.

\bibliographystyle{plainnat}
\bibliography{IEEEabrv,reference}

\begin{thebibliography}{89}
\providecommand{\natexlab}[1]{#1}
\providecommand{\url}[1]{\texttt{#1}}
\expandafter\ifx\csname urlstyle\endcsname\relax
  \providecommand{\doi}[1]{doi: #1}\else
  \providecommand{\doi}{doi: \begingroup \urlstyle{rm}\Url}\fi

\bibitem[Abel et~al.(2013)Abel, Herder, Houben, Henze, and Krause]{no:26}
Fabian Abel, Eelco Herder, Geert-Jan Houben, Nicola Henze, and Daniel Krause.
\newblock Cross-system user modeling and personalization on the social web.
\newblock \emph{User Modeling and User-Adapted Interaction}, 23\penalty0
  (2):\penalty0 169--209, 2013.

\bibitem[Aharon et~al.(2015)Aharon, Hillel, Kagian, Lempel, Makabee, and
  Nissim]{no:57}
Michal Aharon, Eshcar Hillel, Amit Kagian, Ronny Lempel, Hayim Makabee, and Raz
  Nissim.
\newblock Watch-it-next: A contextual tv recommendation system.
\newblock In \emph{Machine Learning and Knowledge Discovery in Databases},
  pages 180--195, Cham, 2015. Springer International Publishing.

\bibitem[Bajaj and Shekhar(2016)]{no:51}
Payal Bajaj and Sumit Shekhar.
\newblock Experience individualization on online tv platforms through
  persona-based account decomposition.
\newblock In \emph{Proceedings of the 24th ACM International Conference on
  Multimedia}, MM '16, pages 252--256, New York, NY, USA, 2016. ACM.

\bibitem[Berkovsky et~al.(2007)Berkovsky, Kuflik, and Ricci]{no:37}
Shlomo Berkovsky, Tsvi Kuflik, and Francesco Ricci.
\newblock Cross-domain mediation in collaborative filtering.
\newblock In \emph{Proceedings of the 11th International Conference on User
  Modeling}, UM '07, pages 355--359, Berlin, Heidelberg, 2007. Springer-Verlag.

\bibitem[Berkovsky et~al.(2008)Berkovsky, Kuflik, and Ricci]{no:27}
Shlomo Berkovsky, Tsvi Kuflik, and Francesco Ricci.
\newblock Mediation of user models for enhanced personalization in recommender
  systems.
\newblock \emph{User Modeling and User-Adapted Interaction}, 18\penalty0
  (3):\penalty0 245--286, 2008.

\bibitem[Bogina and Kuflik(2017)]{no:21}
Veronika Bogina and Tsvi Kuflik.
\newblock Incorporating dwell time in session-based recommendations with
  recurrent neural networks.
\newblock In \emph{Proceedings of RecTemp Workshop co-located with ACM RecSys},
  RecTemp '17, pages 57--59, New York, NY, USA, 2017. ACM.

\bibitem[Cao et~al.(2010)Cao, Liu, and Yang]{no:41}
Bin Cao, Nathan~Nan Liu, and Qiang Yang.
\newblock Transfer learning for collective link prediction in multiple
  heterogenous domains.
\newblock In \emph{Proceedings of the 27th International Conference on
  International Conference on Machine Learning}, ICML'10, pages 159--166, USA,
  2010. Omnipress.

\bibitem[Chen et~al.(2021)Chen, Li, Yan, and Yang]{9319527}
Chao Chen, Dongshen Li, Junchi Yan, and Xiaokang Yang.
\newblock Modeling dynamic user preference via dictionary learning for
  sequential recommendation.
\newblock \emph{IEEE Transactions on Knowledge and Data Engineering}, pages
  1--1, 2021.

\bibitem[Chen et~al.(2017)Chen, Zheng, Gao, Zhou, Zeng, and Chen]{no:44}
Leihui Chen, Jianbing Zheng, Ming Gao, Aoying Zhou, Wei Zeng, and Hui Chen.
\newblock Tlrec: Transfer learning for cross-domain recommendation.
\newblock In \emph{2017 IEEE International Conference on Big Knowledge}, ICBK
  '17, pages 167--172, New York, NY, USA, 2017. IEEE.

\bibitem[Chen et~al.(2012)Chen, Moore, Turnbull, and
  Joachims]{chen2012playlist}
Shuo Chen, Josh~L. Moore, Douglas Turnbull, and Thorsten Joachims.
\newblock Playlist prediction via metric embedding.
\newblock In \emph{Proceedings of the 18th ACM SIGKDD International Conference
  on Knowledge Discovery and Data Mining}, KDD '12, pages 714--722, New York,
  NY, USA, 2012. ACM.

\bibitem[Chen et~al.(2018)Chen, Xu, Zhang, Tang, Cao, Qin, and
  Zha]{Chen:2018:SRU:3159652.3159668}
Xu~Chen, Hongteng Xu, Yongfeng Zhang, Jiaxi Tang, Yixin Cao, Zheng Qin, and
  Hongyuan Zha.
\newblock Sequential recommendation with user memory networks.
\newblock In \emph{Proceedings of the Eleventh ACM International Conference on
  Web Search and Data Mining}, WSDM '18, pages 108--116, New York, NY, USA,
  2018. ACM.

\bibitem[Cheng et~al.(2017)Cheng, Shen, Zhu, Kankanhalli, and
  Nie]{cheng2017exploiting}
Zhiyong Cheng, Jialie Shen, Lei Zhu, Mohan~S Kankanhalli, and Liqiang Nie.
\newblock Exploiting music play sequence for music recommendation.
\newblock In \emph{Proceedings of the 26th International Joint Conference on
  Artificial Intelligence}, IJCAI '17, pages 3654--3660. AAAI Press, 2017.

\bibitem[Cheng et~al.(2018)Cheng, Ding, Zhu, and Kankanhalli]{no:58}
Zhiyong Cheng, Ying Ding, Lei Zhu, and Mohan Kankanhalli.
\newblock Aspect-aware latent factor model: Rating prediction with ratings and
  reviews.
\newblock In \emph{Proceedings of the 2018 World Wide Web Conference}, WWW '18,
  pages 639--648, Republic and Canton of Geneva, Switzerland, 2018.
  International World Wide Web Conferences Steering Committee.

\bibitem[Cremonesi and Quadrana(2014)]{no:43}
Paolo Cremonesi and Massimo Quadrana.
\newblock Cross-domain recommendations without overlapping data: Myth or
  reality?
\newblock In \emph{Proceedings of the 8th ACM Conference on Recommender
  Systems}, RecSys '14, pages 297--300, New York, NY, USA, 2014. ACM.

\bibitem[Cui et~al.(2020)Cui, Wu, Liu, Zhong, and Wang]{8534409}
Qiang Cui, Shu Wu, Qiang Liu, Wen Zhong, and Liang Wang.
\newblock {MV-RNN}: A multi-view recurrent neural network for sequential
  recommendation.
\newblock \emph{IEEE Transactions on Knowledge and Data Engineering},
  32\penalty0 (2):\penalty0 317--331, 2020.

\bibitem[Do et~al.(2021)Do, Liu, Fan, and Tao]{8742537}
Quan Do, Wei Liu, Jin Fan, and Dacheng Tao.
\newblock Unveiling hidden implicit similarities for cross-domain
  recommendation.
\newblock \emph{IEEE Transactions on Knowledge and Data Engineering},
  33\penalty0 (1):\penalty0 302--315, 2021.

\bibitem[Doan and Sahebi(2020)]{9356204}
Thanh-Nam Doan and Shaghayegh Sahebi.
\newblock {TransCrossCF}: Transition-based cross-domain collaborative
  filtering.
\newblock In \emph{2020 19th IEEE International Conference on Machine Learning
  and Applications (ICMLA)}, pages 320--327, 2020.

\bibitem[Donkers et~al.(2017)Donkers, Loepp, and
  Ziegler]{Donkers:2017:SUR:3109859.3109877}
Tim Donkers, Benedikt Loepp, and J\"{u}rgen Ziegler.
\newblock Sequential user-based recurrent neural network recommendations.
\newblock In \emph{Proceedings of the Eleventh ACM Conference on Recommender
  Systems}, RecSys '17, pages 152--160, New York, NY, USA, 2017. ACM.

\bibitem[Elkahky et~al.(2015)Elkahky, Song, and He]{no:30}
Ali~Mamdouh Elkahky, Yang Song, and Xiaodong He.
\newblock A multi-view deep learning approach for cross domain user modeling in
  recommendation systems.
\newblock In \emph{Proceedings of the 24th International Conference on World
  Wide Web}, WWW '15, pages 278--288, Republic and Canton of Geneva,
  Switzerland, 2015. International World Wide Web Conferences Steering
  Committee.

\bibitem[Fang et~al.(2020)Fang, Zhang, Shu, and Guo]{10.1145/3426723}
Hui Fang, Danning Zhang, Yiheng Shu, and Guibing Guo.
\newblock Deep learning for sequential recommendation: Algorithms, influential
  factors, and evaluations.
\newblock \emph{ACM Trans. Inf. Syst.}, 39\penalty0 (1), November 2020.

\bibitem[Farseev et~al.(2017)Farseev, Samborskii, Filchenkov, and Chua]{no:31}
Aleksandr Farseev, Ivan Samborskii, Andrey Filchenkov, and Tat-Seng Chua.
\newblock Cross-domain recommendation via clustering on multi-layer graphs.
\newblock In \emph{Proceedings of the 40th International ACM SIGIR Conference
  on Research and Development in Information Retrieval}, SIGIR '17, pages
  195--204, New York, NY, USA, 2017. ACM.

\bibitem[Fern{\'a}ndez-Tob{\'\i}as et~al.(2012)Fern{\'a}ndez-Tob{\'\i}as,
  Cantador, Kaminskas, and Ricci]{no:32}
Ignacio Fern{\'a}ndez-Tob{\'\i}as, Iv{\'a}n Cantador, Marius Kaminskas, and
  Francesco Ricci.
\newblock Cross-domain recommender systems: A survey of the state of the art.
\newblock In \emph{Proceedings of the 2nd Spanish Conference on Information
  Retrieval}, CERI '12, pages~--, 2012.

\bibitem[Fern\'{a}ndez-Tob\'{\i}as et~al.(2016)Fern\'{a}ndez-Tob\'{\i}as,
  Tomeo, Cantador, Di~Noia, and
  Di~Sciascio]{Fernandez-Tobias:2016:ADC:2959100.2959175}
Ignacio Fern\'{a}ndez-Tob\'{\i}as, Paolo Tomeo, Iv\'{a}n Cantador, Tommaso
  Di~Noia, and Eugenio Di~Sciascio.
\newblock Accuracy and diversity in cross-domain recommendations for cold-start
  users with positive-only feedback.
\newblock In \emph{Proceedings of the 10th ACM Conference on Recommender
  Systems}, RecSys '16, pages 119--122, New York, NY, USA, 2016. ACM.

\bibitem[Forsati et~al.(2014)Forsati, Mahdavi, Shamsfard, and
  Sarwat]{Forsati:2014:MFE:2684820.2641564}
Rana Forsati, Mehrdad Mahdavi, Mehrnoush Shamsfard, and Mohamed Sarwat.
\newblock Matrix factorization with explicit trust and distrust side
  information for improved social recommendation.
\newblock \emph{ACM Transactions on Information Systems}, 32\penalty0
  (4):\penalty0 17:1--17:38, 2014.

\bibitem[Glorot and Bengio(2010)]{glorot2010understanding}
Xavier Glorot and Yoshua Bengio.
\newblock Understanding the difficulty of training deep feedforward neural
  networks.
\newblock In \emph{In Proceedings of the International Conference on Artificial
  Intelligence and Statistics Society for Artificial Intelligence and
  Statistics}, AISTATS '10, pages 249--256, 2010.

\bibitem[He and McAuley(2016{\natexlab{a}})]{He:2016:UDM:2872427.2883037}
Ruining He and Julian McAuley.
\newblock Ups and downs: Modeling the visual evolution of fashion trends with
  one-class collaborative filtering.
\newblock In \emph{Proceedings of the 25th International Conference on World
  Wide Web}, WWW '16, pages 507--517, Republic and Canton of Geneva,
  Switzerland, 2016{\natexlab{a}}. International World Wide Web Conferences
  Steering Committee.

\bibitem[He and McAuley(2016{\natexlab{b}})]{he2016fusing}
Ruining He and Julian McAuley.
\newblock Fusing similarity models with markov chains for sparse sequential
  recommendation.
\newblock In \emph{Proceedings of the IEEE 16th International Conference on
  Data Mining}, ICDM '01, pages 191--200, New York, NY, USA,
  2016{\natexlab{b}}. IEEE.

\bibitem[He et~al.(2017)He, Liao, Zhang, Nie, Hu, and Chua]{no:33}
Xiangnan He, Lizi Liao, Hanwang Zhang, Liqiang Nie, Xia Hu, and Tat-Seng Chua.
\newblock Neural collaborative filtering.
\newblock In \emph{Proceedings of the 26th International Conference on World
  Wide Web}, WWW '17, pages 173--182, Republic and Canton of Geneva,
  Switzerland, 2017. International World Wide Web Conferences Steering
  Committee.

\bibitem[He et~al.(2018)He, He, Du, and Chua]{no:59}
Xiangnan He, Zhankui He, Xiaoyu Du, and Tat-Seng Chua.
\newblock Adversarial personalized ranking for recommendation.
\newblock In \emph{The 41st International ACM SIGIR Conference on Research and
  Development in Information Retrieval}, SIGIR '18, pages 355--364, New York,
  NY, USA, 2018. ACM.

\bibitem[Hidasi et~al.(2016{\natexlab{a}})Hidasi, Karatzoglou, Baltrunas, and
  Tikk]{no:19}
Bal{\'{a}}zs Hidasi, Alexandros Karatzoglou, Linas Baltrunas, and Domonkos
  Tikk.
\newblock Session-based recommendations with recurrent neural networks.
\newblock In \emph{International Conference on Learning Representations}, ICLR
  '16, 2016{\natexlab{a}}.

\bibitem[Hidasi et~al.(2016{\natexlab{b}})Hidasi, Quadrana, Karatzoglou, and
  Tikk]{no:22}
Bal\'{a}zs Hidasi, Massimo Quadrana, Alexandros Karatzoglou, and Domonkos Tikk.
\newblock Parallel recurrent neural network architectures for feature-rich
  session-based recommendations.
\newblock In \emph{Proceedings of the 10th ACM Conference on Recommender
  Systems}, RecSys '16, pages 241--248, New York, NY, USA, 2016{\natexlab{b}}.
  ACM.

\bibitem[Hu et~al.(2018)Hu, Zhang, and Yang]{no:34}
Guangneng Hu, Yu~Zhang, and Qiang Yang.
\newblock Conet: Collaborative cross networks for cross-domain recommendation.
\newblock In \emph{Proceedings of the 27th ACM International Conference on
  Information and Knowledge Management}, CIKM '18, pages 667--676, New York,
  NY, USA, 2018. ACM.

\bibitem[Hu et~al.(2013)Hu, Cao, Xu, Cao, Gu, and Zhu]{no:42}
Liang Hu, Jian Cao, Guandong Xu, Longbing Cao, Zhiping Gu, and Can Zhu.
\newblock Personalized recommendation via cross-domain triadic factorization.
\newblock In \emph{Proceedings of the 22Nd International Conference on World
  Wide Web}, WWW '13, pages 595--606, New York, NY, USA, 2013. ACM.

\bibitem[Huang et~al.(2018)Huang, Zhao, Dou, Wen, and
  Chang]{Huang:2018:ISR:3209978.3210017}
Jin Huang, Wayne~Xin Zhao, Hongjian Dou, Ji-Rong Wen, and Edward~Y. Chang.
\newblock Improving sequential recommendation with knowledge-enhanced memory
  networks.
\newblock In \emph{The 41st International ACM SIGIR Conference on Research and
  Development in Information Retrieval}, SIGIR '18, pages 505--514, New York,
  NY, USA, 2018. ACM.

\bibitem[Huang et~al.(2019)Huang, Ren, Zhao, He, Wen, and
  Dong]{Huang:2019:TMR:3289600.3290972}
Jin Huang, Zhaochun Ren, Wayne~Xin Zhao, Gaole He, Ji-Rong Wen, and Daxiang
  Dong.
\newblock Taxonomy-aware multi-hop reasoning networks for sequential
  recommendation.
\newblock In \emph{Proceedings of the Twelfth ACM International Conference on
  Web Search and Data Mining}, WSDM '19, pages 573--581, New York, NY, USA,
  2019. ACM.

\bibitem[Huang et~al.(2013)Huang, He, Gao, Deng, Acero, and
  Heck]{huang2013learning}
Po-Sen Huang, Xiaodong He, Jianfeng Gao, Li~Deng, Alex Acero, and Larry Heck.
\newblock Learning deep structured semantic models for web search using
  clickthrough data.
\newblock In \emph{Proceedings of the 22Nd ACM International Conference on
  Information \& Knowledge Management}, CIKM '13, pages 2333--2338, New York,
  NY, USA, 2013. ACM.

\bibitem[Jannach and Ludewig(2017)]{no:24}
Dietmar Jannach and Malte Ludewig.
\newblock When recurrent neural networks meet the neighborhood for
  session-based recommendation.
\newblock In \emph{Proceedings of the 11th ACM Conference on Recommender
  Systems}, RecSys '17, pages 306--310, New York, NY, USA, 2017. ACM.

\bibitem[Jiang et~al.(2018)Jiang, Li, Chen, and Wang]{no:49}
Jyun-Yu Jiang, Cheng-Te Li, Yian Chen, and Wei Wang.
\newblock Identifying users behind shared accounts in online streaming
  services.
\newblock In \emph{The 41st International ACM SIGIR Conference on Research
  \&\#38; Development in Information Retrieval}, SIGIR '18, pages 65--74, New
  York, NY, USA, 2018. ACM.

\bibitem[Kanagawa et~al.(2018)Kanagawa, Kobayashi, Shimizu, Tagami, and
  Suzuki]{no:46}
Heishiro Kanagawa, Hayato Kobayashi, Nobuyuki Shimizu, Yukihiro Tagami, and
  Taiji Suzuki.
\newblock Cross-domain recommendation via deep domain adaptation.
\newblock \emph{arXiv preprint arXiv:1803.03018}, 2018.

\bibitem[Koren et~al.(2009)Koren, Bell, and
  Volinsky]{Koren:2009:MFT:1608565.1608614}
Yehuda Koren, Robert Bell, and Chris Volinsky.
\newblock Matrix factorization techniques for recommender systems.
\newblock \emph{Computer}, 42\penalty0 (8):\penalty0 30--37, 2009.

\bibitem[Lesaege et~al.(2016)Lesaege, Schnitzler, Lambert, and
  Vigouroux]{7837939}
Cl{\'e}ment Lesaege, Fran{\c c}ois Schnitzler, Anne Lambert, and Jean-Ronan
  Vigouroux.
\newblock Time-aware user identification with topic models.
\newblock In \emph{2016 IEEE 16th International Conference on Data Mining},
  ICDM '16, pages 997--1002, New York, NY, USA, 2016. IEEE.

\bibitem[Li et~al.(2009)Li, Yang, and Xue]{no:28}
Bin Li, Qiang Yang, and Xiangyang Xue.
\newblock Can movies and books collaborate? cross-domain collaborative
  filtering for sparsity reduction.
\newblock In \emph{Proceedings of the 17th International Joint Conference on
  Artificial Intelligence}, IJCAI '09, pages 2052--2057. AAAI Press, 2009.

\bibitem[Li et~al.(2021)Li, Wang, Zhang, Ma, Cui, and Zhu]{9319534}
Haoyang Li, Xin Wang, Ziwei Zhang, Jianxin Ma, Peng Cui, and Wenwu Zhu.
\newblock Intention-aware sequential recommendation with structured intent
  transition.
\newblock \emph{IEEE Transactions on Knowledge and Data Engineering}, pages
  1--1, 2021.

\bibitem[Li et~al.(2017)Li, Ren, Chen, Ren, Lian, and Ma]{no:25}
Jing Li, Pengjie Ren, Zhumin Chen, Zhaochun Ren, Tao Lian, and Jun Ma.
\newblock Neural attentive session-based recommendation.
\newblock In \emph{Proceedings of the 2017 ACM on Conference on Information and
  Knowledge Management}, CIKM '17, pages 1419--1428, New York, NY, USA, 2017.
  ACM.

\bibitem[Lian et~al.(2017)Lian, Zhang, Xie, and Sun]{no:36}
Jianxun Lian, Fuzheng Zhang, Xing Xie, and Guangzhong Sun.
\newblock Cccfnet: A content-boosted collaborative filtering neural network for
  cross domain recommender systems.
\newblock In \emph{Proceedings of the 26th International Conference on World
  Wide Web Companion}, WWW '17 Companion, pages 817--818, Republic and Canton
  of Geneva, Switzerland, 2017. International World Wide Web Conferences
  Steering Committee.

\bibitem[Linden et~al.(2003)Linden, Smith, and York]{no:3}
Greg Linden, Brent Smith, and Jeremy York.
\newblock Amazon.com recommendations: Item-to-item collaborative filtering.
\newblock \emph{IEEE Internet Computing}, 7\penalty0 (1):\penalty0 76--80,
  2003.

\bibitem[Liu et~al.(2016)Liu, Wu, Wang, Li, and Wang]{7837948}
Qiang Liu, Shu Wu, Diyi Wang, Zhaokang Li, and Liang Wang.
\newblock Context-aware sequential recommendation.
\newblock In \emph{Proceedings of the IEEE 16th International Conference on
  Data Mining}, ICDM '16, pages 1053--1058, New York, NY, USA, 2016. IEEE.

\bibitem[Liu et~al.(2020)Liu, Tian, Zhao, and Zhang]{9346486}
Zhen Liu, Jingyu Tian, Lingxi Zhao, and Yanling Zhang.
\newblock Attentive-feature transfer based on mapping for cross-domain
  recommendation.
\newblock In \emph{2020 International Conference on Data Mining Workshops
  (ICDMW)}, pages 151--158, 2020.

\bibitem[Loni et~al.(2014)Loni, Shi, Larson, and Hanjalic]{no:38}
Babak Loni, Yue Shi, Martha Larson, and Alan Hanjalic.
\newblock Cross-domain collaborative filtering with factorization machines.
\newblock In \emph{Proceedings of the 4th Spanish Conference on Information
  Retrieval}, CERI '14, pages 656--661, 2014.

\bibitem[Lu et~al.(2013)Lu, Zhong, Zhao, Xiang, Pan, and
  Yang]{doi:10.1137/1.9781611972832.71}
Zhongqi Lu, Erheng Zhong, Lili Zhao, Evan~Wei Xiang, Weike Pan, and Qiang Yang.
\newblock Selective transfer learning for cross domain recommendation.
\newblock In \emph{Proceedings of the 2013 SIAM International Conference on
  Data Mining}, SIAM '13, pages 641--649, Austin, Texas, USA, 2013.

\bibitem[Lucic et~al.(2021)Lucic, Srikumar, Bhatt, Xiang, Taly, Liao, and
  de~Rijke]{lucic-2021-multistakeholder}
Ana Lucic, Madhulika Srikumar, Umang Bhatt, Alice Xiang, Ankur Taly, Q.~Vera
  Liao, and Maarten de~Rijke.
\newblock A multistakeholder approach towards evaluating ai transparency
  mechanisms.
\newblock In \emph{ACM CHI Workshop on Operationalizing Human-Centered
  Perspectives in Explainable AI}. ACM, May 2021.

\bibitem[Ma et~al.(2019)Ma, Ren, Lin, Chen, Ma, and de~Rijke]{ma-2019-pi-net}
Muyang Ma, Pengjie Ren, Yujie Lin, Zhumin Chen, Jun Ma, and Maarten de~Rijke.
\newblock $\pi$-net: A parallel information-sharing network for cross-domain
  shared-account sequential recommendations.
\newblock In \emph{The 42st International ACM SIGIR Conference on Research and
  Development in Information Retrieval}, SIGIR '19, pages 685--694, New York,
  NY, USA, 2019. ACM.

\bibitem[Ma et~al.(2018)Ma, Zhang, Wang, Cui, and
  Huang]{Ma:2018:MRM:3209978.3210026}
Renfeng Ma, Qi~Zhang, Jiawen Wang, Lizhen Cui, and Xuanjing Huang.
\newblock Mention recommendation for multimodal microblog with cross-attention
  memory network.
\newblock In \emph{The 41st International ACM SIGIR Conference on Research and
  Development in Information Retrieval}, SIGIR '18, pages 195--204, New York,
  NY, USA, 2018. ACM.

\bibitem[Mei et~al.(2018)Mei, Ren, Chen, Nie, Ma, and Nie]{no:60}
Lei Mei, Pengjie Ren, Zhumin Chen, Liqiang Nie, Jun Ma, and Jian-Yun Nie.
\newblock An attentive interaction network for context-aware recommendations.
\newblock In \emph{Proceedings of the 27th ACM International Conference on
  Information and Knowledge Management}, CIKM '18, pages 157--166, New York,
  NY, USA, 2018. ACM.

\bibitem[Misra et~al.(2016)Misra, Shrivastava, Gupta, and
  Hebert]{misra2016cross}
Ishan Misra, Abhinav Shrivastava, Abhinav Gupta, and Martial Hebert.
\newblock Cross-stitch networks for multi-task learning.
\newblock In \emph{The IEEE Conference on Computer Vision and Pattern
  Recognition}, CVPR '16, pages 3994--4003, New York, NY, USA, 2016. IEEE.

\bibitem[Onal et~al.(2018)Onal, Zhang, Altingovde, Rahman, Karagoz, Braylan,
  Dang, Chang, Kim, McNamara, Angert, Banner, Khetan, McDonnell, Nguyen, Xu,
  Wallace, de~Rijke, and Lease]{onal-neural-2018}
Kezban~Dilek Onal, Ye~Zhang, Ismail~Sengor Altingovde, Md~Mustafizur Rahman,
  Pinar Karagoz, Alex Braylan, Brandon Dang, Heng-Lu Chang, Henna Kim, Quinten
  McNamara, Aaron Angert, Edward Banner, Vivek Khetan, Tyler McDonnell,
  An~Thanh Nguyen, Dan Xu, Byron~C. Wallace, Maarten de~Rijke, and Matthew
  Lease.
\newblock Neural information retrieval: At the end of the early years.
\newblock \emph{Information Retrieval Journal}, 21\penalty0 (2--3):\penalty0
  111--182, June 2018.

\bibitem[Pan et~al.(2010)Pan, Xiang, Liu, and Yang]{no:29}
Weike Pan, Evan~Wei Xiang, Nathan~Nan Liu, and Qiang Yang.
\newblock Transfer learning in collaborative filtering for sparsity reduction.
\newblock In \emph{The 24th {AAAI} Conference on Artificial Intelligence}, AAAI
  '10, pages 230--235. {AAAI} Press, 2010.

\bibitem[Pascanu et~al.(2013)Pascanu, Mikolov, and
  Bengio]{pascanu2013difficulty}
Razvan Pascanu, Tomas Mikolov, and Yoshua Bengio.
\newblock On the difficulty of training recurrent neural networks.
\newblock In \emph{Proceedings of the 30th International Conference on
  International Conference on Machine Learning}, ICML '13, pages
  III--1310--III--1318. JMLR.org, 2013.

\bibitem[Quadrana et~al.(2017)Quadrana, Karatzoglou, Hidasi, and
  Cremonesi]{no:1}
Massimo Quadrana, Alexandros Karatzoglou, Balzs Hidasi, and Paolo Cremonesi.
\newblock Personalizing session-based recommendations with hierarchical
  recurrent neural networks.
\newblock In \emph{Proceedings of the 11th ACM Conference on Recommender
  Systems}, RecSys '17, pages 130--137, New York, NY, USA, 2017. ACM.

\bibitem[Quadrana et~al.(2018)Quadrana, Cremonesi, and
  Jannach]{Quadrana:2018:SRS:3236632.3190616}
Massimo Quadrana, Paolo Cremonesi, and Dietmar Jannach.
\newblock Sequence-aware recommender systems.
\newblock \emph{ACM Computing Surveys}, 51\penalty0 (4):\penalty0 66:1--66:36,
  2018.

\bibitem[Ren et~al.(2019)Ren, Chen, Li, Ren, Ma, and de~Rijke]{no:48}
Pengjie Ren, Zhumin Chen, Jing Li, Zhaochun Ren, Jun Ma, and Maarten de~Rijke.
\newblock Repeatnet: A repeat aware neural recommendation machine for
  session-based recommendation.
\newblock In \emph{The Thirty-Third {AAAI} Conference on Artificial
  Intelligence}, AAAI '19, pages 4806--4813. {AAAI} Press, 2019.

\bibitem[Rendle et~al.(2009)Rendle, Freudenthaler, Gantner, and
  Schmidt-Thieme]{no:61}
Steffen Rendle, Christoph Freudenthaler, Zeno Gantner, and Lars Schmidt-Thieme.
\newblock Bpr: Bayesian personalized ranking from implicit feedback.
\newblock In \emph{Proceedings of the Twenty-Fifth Conference on Uncertainty in
  Artificial Intelligence}, UAI '09, pages 452--461, Arlington, Virginia,
  United States, 2009. AUAI Press.

\bibitem[Rendle et~al.(2010)Rendle, Freudenthaler, and
  Schmidt-Thieme]{Rendle:2010:FPM:1772690.1772773}
Steffen Rendle, Christoph Freudenthaler, and Lars Schmidt-Thieme.
\newblock Factorizing personalized markov chains for next-basket
  recommendation.
\newblock In \emph{Proceedings of the 19th International Conference on World
  Wide Web}, WWW '10, pages 811--820, New York, NY, USA, 2010. ACM.

\bibitem[Sarwar et~al.(2001)Sarwar, Karypis, Konstan, and
  Riedl]{Sarwar:2001:ICF:371920.372071}
Badrul Sarwar, George Karypis, Joseph Konstan, and John Riedl.
\newblock Item-based collaborative filtering recommendation algorithms.
\newblock In \emph{Proceedings of the 10th International Conference on World
  Wide Web}, WWW '01, pages 285--295, New York, NY, USA, 2001. ACM.

\bibitem[Shani et~al.(2005)Shani, Heckerman, and Brafman]{no:4}
Guy Shani, David Heckerman, and Ronen~I. Brafman.
\newblock An mdp-based recommender system.
\newblock \emph{Journal of Machine Learning Research}, 6:\penalty0 1265--1295,
  2005.

\bibitem[Shapira et~al.(2013)Shapira, Rokach, and Freilikhman]{no:39}
Bracha Shapira, Lior Rokach, and Shirley Freilikhman.
\newblock Facebook single and cross domain data for recommendation systems.
\newblock \emph{User Modeling and User-Adapted Interaction}, 23\penalty0
  (2):\penalty0 211--247, 2013.

\bibitem[Srivastava et~al.(2014)Srivastava, Hinton, Krizhevsky, Sutskever, and
  Salakhutdinov]{srivastava2014dropout}
Nitish Srivastava, Geoffrey Hinton, Alex Krizhevsky, Ilya Sutskever, and Ruslan
  Salakhutdinov.
\newblock Dropout: A simple way to prevent neural networks from overfitting.
\newblock \emph{Journal of Machine Learning Research}, 15\penalty0
  (1):\penalty0 1929--1958, 2014.

\bibitem[Tan et~al.(2021)Tan, Zhang, Liu, Huang, Yang, Zhou, and Hu]{unknown}
Qiaoyu Tan, Jianwei Zhang, Ninghao Liu, Xiao Huang, Hongxia Yang, Jingren Zhou,
  and Xia Hu.
\newblock Dynamic memory based attention network for sequential recommendation,
  02 2021.

\bibitem[Tan et~al.(2016)Tan, Xu, and Liu]{no:20}
Yong~Kiam Tan, Xinxing Xu, and Yong Liu.
\newblock Improved recurrent neural networks for session-based recommendations.
\newblock In \emph{Proceedings of the 1st Workshop on Deep Learning for
  Recommender Systems}, DLRS '16, pages 17--22, New York, NY, USA, 2016. ACM.

\bibitem[Tang and Wang(2018)]{Tang:2018:PTS:3159652.3159656}
Jiaxi Tang and Ke~Wang.
\newblock Personalized top-n sequential recommendation via convolutional
  sequence embedding.
\newblock In \emph{Proceedings of the Eleventh ACM International Conference on
  Web Search and Data Mining}, WSDM '18, pages 565--573, New York, NY, USA,
  2018. ACM.

\bibitem[Tang et~al.(2012)Tang, Wu, Sun, and Su]{Tang:2012:CCR:2339530.2339730}
Jie Tang, Sen Wu, Jimeng Sun, and Hang Su.
\newblock Cross-domain collaboration recommendation.
\newblock In \emph{Proceedings of the 18th ACM SIGKDD International Conference
  on Knowledge Discovery and Data Mining}, KDD '12, pages 1285--1293, New York,
  NY, USA, 2012. ACM.

\bibitem[Vasile et~al.(2016)Vasile, Smirnova, and
  Conneau]{Vasile:2016:MPE:2959100.2959160}
Flavian Vasile, Elena Smirnova, and Alexis Conneau.
\newblock Meta-prod2vec: Product embeddings using side-information for
  recommendation.
\newblock In \emph{Proceedings of the 10th ACM Conference on Recommender
  Systems}, RecSys '16, pages 225--232, New York, NY, USA, 2016. ACM.

\bibitem[Verstrepen and Goethals(2015)]{no:52}
Koen Verstrepen and Bart Goethals.
\newblock Top-n recommendation for shared accounts.
\newblock In \emph{Proceedings of the 9th ACM Conference on Recommender
  Systems}, RecSys '15, pages 59--66, New York, NY, USA, 2015. ACM.

\bibitem[Wang et~al.(2019)Wang, Ren, Mei, Chen, Jun, and
  de~Rijke]{Wang:3209978.3210026}
Meirui Wang, Pengjie Ren, Lei Mei, Zhumin Chen, Ma~Jun, and Maarten de~Rijke.
\newblock A collaborative session-based recommendation approach with parallel
  memory modules.
\newblock In \emph{The 42st International ACM SIGIR Conference on Research and
  Development in Information Retrieval}, SIGIR '19, pages 345--354, New York,
  NY, USA, 2019. ACM.

\bibitem[Wang et~al.(2015)Wang, Guo, Lan, Xu, Wan, and Cheng]{no:16}
Pengfei Wang, Jiafeng Guo, Yanyan Lan, Jun Xu, Shengxian Wan, and Xueqi Cheng.
\newblock Learning hierarchical representation model for next basket
  recommendation.
\newblock In \emph{Proceedings of the 38th International ACM SIGIR Conference
  on Research and Development in Information Retrieval}, SIGIR '15, pages
  403--412, New York, NY, USA, 2015. ACM.

\bibitem[Wang et~al.(2017)Wang, He, Nie, and Chua]{no:35}
Xiang Wang, Xiangnan He, Liqiang Nie, and Tat-Seng Chua.
\newblock Item silk road: Recommending items from information domains to social
  users.
\newblock In \emph{Proceedings of the 40th International ACM SIGIR Conference
  on Research and Development in Information Retrieval}, SIGIR '17, pages
  185--194, New York, NY, USA, 2017. ACM.

\bibitem[Wang et~al.(2014)Wang, Yang, He, and Gu]{no:53}
Zhijin Wang, Yan Yang, Liang He, and Junzhong Gu.
\newblock User identification within a shared account: Improving ip-tv
  recommender performance.
\newblock In \emph{Advances in Databases and Information Systems}, ADBIS '14,
  pages 219--233, Cham, 2014. Springer International Publishing.

\bibitem[Wu et~al.(2013)Wu, Liu, Chen, He, Lv, Cao, and
  Hu]{Wu:2013:PNR:2507157.2507215}
Xiang Wu, Qi~Liu, Enhong Chen, Liang He, Jingsong Lv, Can Cao, and Guoping Hu.
\newblock Personalized next-song recommendation in online karaokes.
\newblock In \emph{Proceedings of the 7th ACM Conference on Recommender
  Systems}, RecSys '13, pages 137--140, New York, NY, USA, 2013. ACM.

\bibitem[Yan et~al.(2020)Yan, Tan, Tsang, Yang, Shi, and Zhang]{8540438}
Yan Yan, Mingkui Tan, Ivor~W. Tsang, Yi~Yang, Qinfeng Shi, and Chengqi Zhang.
\newblock Fast and low memory cost matrix factorization: Algorithm, analysis,
  and case study.
\newblock \emph{IEEE Transactions on Knowledge and Data Engineering},
  32\penalty0 (2):\penalty0 288--301, 2020.

\bibitem[Yang et~al.(2017)Yang, Sarkhel, Mitra, and
  Swaminathan]{yang2017personalized}
Shuo Yang, Somdeb Sarkhel, Saayan Mitra, and Viswanathan Swaminathan.
\newblock Personalized video recommendations for shared accounts.
\newblock In \emph{2017 IEEE International Symposium on Multimedia}, ISM '17,
  pages 256--259, 2017.

\bibitem[Yang et~al.(2015)Yang, Hu, He, Ni, and Wang]{no:54}
Yan Yang, Qinmin Hu, Liang He, Minjie Ni, and Zhijin Wang.
\newblock Adaptive temporal model for iptv recommendation.
\newblock In \emph{Web-Age Information Management}, WAIM '15, pages 260--271,
  Cham, 2015. Springer International Publishing.

\bibitem[Yap et~al.(2012)Yap, Li, and Philip]{yap2012effective}
Ghim-Eng Yap, Xiao-Li Li, and S~Yu Philip.
\newblock Effective next-items recommendation via personalized sequential
  pattern mining.
\newblock In \emph{Database Systems for Advanced Applications}, DASFAA '12,
  pages 48--64, Berlin, Heidelberg, 2012. Springer Berlin Heidelberg.

\bibitem[Yu et~al.(2016)Yu, Liu, Wu, Wang, and Tan]{no:18}
Feng Yu, Qiang Liu, Shu Wu, Liang Wang, and Tieniu Tan.
\newblock A dynamic recurrent model for next basket recommendation.
\newblock In \emph{Proceedings of the 39th International ACM SIGIR Conference
  on Research and Development in Information Retrieval}, SIGIR '16, pages
  729--732, New York, NY, USA, 2016. ACM.

\bibitem[Zhang et~al.(2012)Zhang, Fawaz, Ioannidis, and Montanari]{no:55}
Amy Zhang, Nadia Fawaz, Stratis Ioannidis, and Andrea Montanari.
\newblock Guess who rated this movie: Identifying users through subspace
  clustering.
\newblock In \emph{Proceedings of the Twenty-Eighth Conference on Uncertainty
  in Artificial Intelligence}, UAI '12, pages 944--953, Arlington, Virginia,
  United States, 2012. AUAI Press.

\bibitem[Zhang et~al.(2021)Zhang, Kong, and Zhang]{9360540}
Hongwei Zhang, Xiangwei Kong, and Yujia Zhang.
\newblock Selective knowledge transfer for cross-domain collaborative
  recommendation.
\newblock \emph{IEEE Access}, pages 1--1, 2021.

\bibitem[Zhao et~al.(2016)Zhao, Cao, and Tan]{no:56}
Yafeng Zhao, Jian Cao, and Yudong Tan.
\newblock Passenger prediction in shared accounts for flight service
  recommendation.
\newblock In \emph{The 10th International Conference on Asia-Pacific Services
  Computing}, APSCC '16, pages 159--172, Cham, 2016. Springer International
  Publishing.

\bibitem[Zhuang et~al.(2010)Zhuang, Luo, Xiong, Xiong, He, and Shi]{no:40}
Fuzhen Zhuang, Ping Luo, Hui Xiong, Yuhong Xiong, Qing He, and Zhongzhi Shi.
\newblock Cross-domain learning from multiple sources: A consensus
  regularization perspective.
\newblock \emph{Transactions on Knowledge and Data Engineering}, 22:\penalty0
  1664--1678, 2010.

\bibitem[Zhuang et~al.(2017)Zhuang, Zhou, Zhang, Ao, Xie, and He]{no:50}
Fuzhen Zhuang, Yingmin Zhou, Fuzheng Zhang, Xiang Ao, Xing Xie, and Qing He.
\newblock Sequential transfer learning: Cross-domain novelty seeking trait
  mining for recommendation.
\newblock In \emph{Proceedings of the 26th International Conference on World
  Wide Web Companion}, WWW '17 Companion, pages 881--882, Republic and Canton
  of Geneva, Switzerland, 2017. International World Wide Web Conferences
  Steering Committee.

\bibitem[Zimdars et~al.(2001)Zimdars, Chickering, and Meek]{zimdars2001using}
Andrew Zimdars, David~Maxwell Chickering, and Christopher Meek.
\newblock Using temporal data for making recommendations.
\newblock In \emph{Proceedings of the Seventeenth Conference on Uncertainty in
  Artificial Intelligence}, UAI '01, pages 580--588, San Francisco, CA, USA,
  2001. Morgan Kaufmann Publishers Inc.

\end{thebibliography}

\end{document}